\documentclass[aps,prb,twocolumn,a4paper,showpacs,superscriptaddress,floatfix,10pt]{revtex4-1}

\usepackage{graphicx}
\usepackage{amsmath}
\usepackage{epsfig}
\usepackage{helvet}
\usepackage{amssymb}
\usepackage{epstopdf}
\usepackage{bbm}
\usepackage{bm}

\def\U{ \mathcal U}
\def\V{ \mathcal V}

\begin{document}

\title{Representation and design of wavelets using unitary circuits}

\author{Glen Evenbly}
\affiliation{Institute for Quantum Information and Matter,
California Institute of Technology, MC 305-16, Pasadena CA 91125, USA}
\affiliation{Department of Physics and Astronomy, University of California, Irvine, CA 92697-4575 USA}
\author{Steven R. White}
\affiliation{Department of Physics and Astronomy, University of California, Irvine, CA 92697-4575 USA}
\date{\today}

\begin{abstract}
The representation of discrete, compact wavelet transformations (WTs) as circuits of local unitary gates is discussed. We employ a similar formalism as used in the multi-scale representation of quantum many-body wavefunctions using unitary circuits, further cementing the relation established in Refs. \onlinecite{MattWavelet,waveMERA} between classical and quantum multi-scale methods. An algorithm for constructing the circuit representation of known orthogonal, dyadic, discrete WTs is presented, and the explicit representation for Daubechies wavelets, coiflets, and symlets is provided. Furthermore, we demonstrate the usefulness of the circuit formalism in designing novel WTs, including various classes of symmetric wavelets and multi-wavelets, boundary wavelets and biorthogonal wavelets.
\end{abstract}


\maketitle
\tableofcontents

\section{Introduction}
The development of compact, orthogonal wavelets\cite{Daub1,Daub2,Daub3,Daub4} represents one of the most significant advances in signal processing from recent times\cite{WaveBook1,WaveBook2,WaveBook3}. Like well established methods based on Fourier analysis, wavelets can be employed to obtain a compressed representation of spatially or temporally correlated data, and thus have found significant applications in image, audio and video compression \cite{WaveImageApp,SigBook,JPEG}. By providing a multi-resolution analysis, a WT can capture correlations over many different length/time scales, which is key to their usefulness in practical applications.

Similarly, multi-scale methods, in the context of the renormalization group (RG) \cite{RGreview}, which provides a systematic framework to implement changes of scale in many body systems, have also had a long and successful history in the context of condensed matter physics and quantum field theories. A recent advance in RG methods for quantum many-body systems is the multi-scale entanglement renormalization ansatz (MERA)\cite{ER,MERA}, where quantum many-body wavefunctions are represented as circuits comprised of local unitary gates. By properly encoding correlations at all length scales, MERA have been demonstrated to be particularly useful in the numerical investigation of quantum many-body systems where many scales of length are important, such as in systems at the critical point of a phase transition\cite{MERAlocal,MERAnonlocal,MERAbook,Alg2,MERAapp1,MERAapp2,MERAapp3,MERAapp4}.

While the conceptual links between WTs and the RG have been previously examined\cite{Battle}, more recently it was shown that a WT acting in the space of fermionic mode operators corresponds precisely to a MERA\cite{MattWavelet,waveMERA}. This wavelet/MERA connection proved useful in allowing construction of the first known analytic MERA for a critical system, and may further prove useful in understanding e.g. the scaling of errors in MERA. The purpose of this paper is to take the connection the other way, and examine whether the tools and ideas developed in the context of the MERA are useful in the understanding and design of WTs. Specifically, we discuss the use of unitary circuits for the representation of standard families of dyadic wavelets (including Daubechies wavelets, coiflets and symlets), and then explore use of unitary circuits for the design of novel wavelet families (including symmetric wavelets of dilation factor $m = 3$ and $m = 4$, symmetric multiwavelets, boundary wavelets, and a new family of symmetric biorthogonal wavelets).   

This paper is organized as follows. Firstly in Sect. \ref{sect:circuit} we introduce our notion of multi-scale unitary circuits, which follow as the classical analogue of MERA quantum circuits. Then in Sect. \ref{sect:wavecircuit} we discuss the representation of discrete, orthogonal, dyadic wavelet transformations as unitary circuits. This includes providing the explicit form of the circuit representation for Daubechies wavelets, coiflets and symlets, as well as an algorithm for obtaining the circuit representation of arbitrary dyadic wavelets. In Sect. \ref{sect:novel} we explore the use of generalized unitary circuits in the design of novel WTs, including, in Sect. \ref{sect:dilation3} families of symmetric dilation factor $m = 3$ wavelets, in Sect. \ref{sect:multi} a family of symmetric multiwavelets, in Sect. \ref{sect:dilation4} a family of symmetric dilation factor $m = 4$ wavelets, in Sect. \ref{sect:boundary} a construction of boundary wavelets, and in Sect. \ref{sect:biorthog} a family of symmetric biorthogonal wavelets. Finally, conclusions and discussion are presented in Sect. \ref{sect:conclusion}.

\section{Unitary Circuits} \label{sect:circuit}
\begin{figure} [!t]
  \begin{centering}
\includegraphics[width=8.5cm]{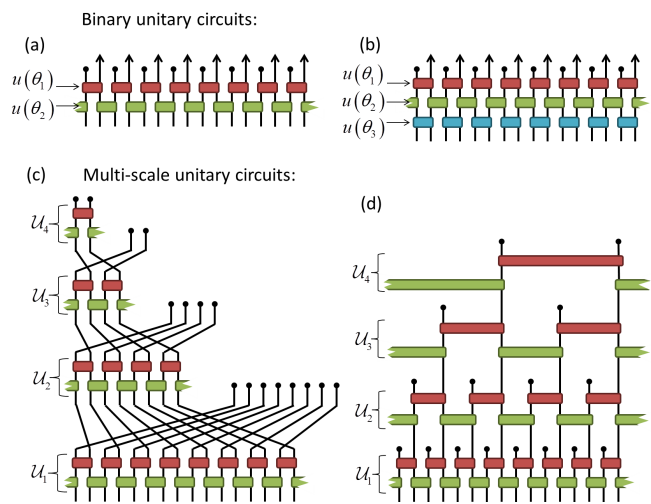}  \end{centering}
\caption{(a) Depiction of a depth $N=2$ binary unitary circuit, built from $2 \times 2$ unitary gates $u$, see Eq. \ref{eq:s1e3}, with angles $\{ \theta_1, \theta_2 \}$. (b) Depth $N=3$ binary unitary circuit parameterized by angles $\{ \theta_1, \theta_2, \theta_3 \}$. (c) Multi-scale unitary circuit formed from composition of $N=2$ binary unitary circuits, see also Eq. \ref{eq:s1e4}. (d) Alternative representation of the same multi-scale circuit from (c).}
  \label{fig:BinCircuit}
\end{figure}

In this section we define our notions of binary and multi-scale unitary circuits, with are the analogue of the MERA circuits\cite{MERA} used for representing quantum many-body wavefunctions. Let $\U$ be a $2M \times 2M$ unitary matrix for integer $M$. We say $\U$ is a binary unitary circuit of depth $N$ if it can be written as a product of matrices $U_{k}$,
\begin{equation}
\U = U_{N}\times U_{N-1}\times\ldots U_{2} \times U_{1}. \label{eq:s1e1}
\end{equation}
where each $U_{k}$ is a $2M \times 2M$ unitary matrix that is the direct sum of $M$ neighboring $2\times 2$ unitary matrices $u(\theta_k)$ (alternating between odd-even and even-odd placement with each layer), i.e. such that
\begin{equation}
{U_k} \equiv \mathop  \bigoplus \limits_{r \textrm{ odd}} {u_{[r,r+1]}}(\theta_k), \label{eq:s1e2}
\end{equation}
for $k$ odd, and with the direct sum over even $r$ sites for $k$ even, see also Fig. \ref{fig:BinCircuit}(a-b) for a diagrammatic representation. The free parameter $\theta_k\in (-\pi,\pi]$ defines an angle of rotation in the unitary matrix $u$,  
\begin{align} 
{u(\theta_k)} & \equiv \left[ {\begin{array}{*{20}{c}}
  {\cos \left( {{\theta _k}} \right)}&{\sin \left( {{\theta _k}} \right)} \\ 
  { - \sin \left( {{\theta _k}} \right)}&{\cos \left( {{\theta _k}} \right)} 
\end{array}} \right],  \nonumber \\ 
& = \cos\left( {{\theta _k}} \right) \left[ {\begin{array}{*{20}{c}}
  1&t_k \\ 
  { -t_k}&{1} 
\end{array}} \right], \label{eq:s1e3}
\end{align}
where $t_k$ is short for $\tan(\theta_k)$. 

A multi-scale circuit, representing a matrix of dimension $2^T \times 2^T$ for positive integer $T$, can be formed from composition of $T$ binary unitary circuits,
\begin{equation} 
\U_1 \circ \U_2 \circ \ldots \circ \U_T. \label{eq:s1e4}
\end{equation} 
Here each $\U_k$ is a binary unitary circuit of linear dimension $2^{T-k+1}$ and the composition is such that $\U_k$ acts only on the even index sites of the previous $\U_{k-1}$, see also Fig. \ref{fig:BinCircuit}(c-d).

%%%%%%%%%%%%%%%%%%%%%%%%%%%%%%%%%%%%%%%%%%%%%%%%%%%%%%
%%%%%%%%%%%%%%%%%%%%%%%%%%%%%%%%%%%%%%%%%%%%%%%%%%%%%%

\section{Wavelets and unitary circuits} \label{sect:wavecircuit}
In this section we describe how discrete WTs can be represented as unitary circuits, first in Sect. \ref{sect:basic} recalling basic properties of wavelets \cite{WaveBook2}, before next in Sect. \ref{sect:D4} providing a specific example of the circuit representation of the D4 Daubechies wavelet, and finally in Sect. \ref{sect:arbitrary} providing an algorithm for constructing the circuit representation of arbitrary dyadic orthogonal wavelets.

\subsection{Basic properties of orthogonal wavelets} \label{sect:basic}
Here we recall the fundamentals of compact, dyadic (i.e. dilation factor $m = 2$), orthogonal wavelets. We seek to construct a set of wavelet functions $\psi_z$ for $z = [1, 2, \ldots]$ on a $1D$ lattice of sites, where $z$ denotes the scale over which the function has non-zero support, such that wavelet $\psi_z$ has a compact support of $O(2^z)$ sites. Additionally, we require that wavelets at different scales should be mutually orthogonal, and wavelets at the same scale should be orthogonal under translations of $x = n 2^z$ sites for integer $n$. The construction of wavelets that fulfill these properties is facilitated by first introducing the set of \emph{scaling functions}, which we denote $\phi_z$ for $z = [1, 2, \ldots]$, where $z$ again denotes scale. The scaling functions are defined recursively through the \emph{refinement equation} \cite{WaveBook3},
\begin{equation}
\phi_{z+1} (x) = \sum\limits_{r = 1}^{2N} {{h_r}\phi_z (2x - r)}, \label{eq:s2e1}
\end{equation}
where ${\bm h} = [h_1, h_2, \ldots, h_{2N}]$ is called the \emph{scaling sequence}, with $N$ the order of the wavelet transform. Note that the scaling function at smallest scale $z=1$ is defined directly from the scaling sequence, $\phi_1(x) \equiv h_r \delta_{x,r}$, with $\delta_{i,j}$ the Kronecker delta function. A necessary condition for a scaling sequence $\bm{h}$ to generate orthogonal wavelets is that it is orthogonal to itself under shifts by an even number of coefficients, 
\begin{equation}
\sum\limits_{r = 1}^{2N - 2m} {{h_r}{h_{r + 2m}} = {\delta _{m,0}}}. \label{eq:s2e2}
\end{equation}
For dyadic wavelets, the wavelet sequence $\bm{g} = [g_1, g_2, \ldots, g_{2N}]$ is defined from the scaling sequence $\bm{h}$ by reversing the order of elements and introducing an alternating minus sign,
\begin{equation}
{g_r} = {\left( { - 1} \right)^{r + 1}}{h_{2N - r + 1}}. \label{eq:s2e3}
\end{equation}
Notice that the wavelet sequence $\bm{g}$ is, by construction, orthogonal to the scaling sequence $\bm{h}$ from which it is derived. The wavelets $\psi_{z}$ can now be defined from the scaling functions; specifically the wavelet $\psi_{z+1}$ at scale $z+1$ is given from linear combinations of translations of scaling functions $\phi_z$ at the previous scale, 
\begin{equation}
\psi_{z+1} (x) = \sum\limits_{r = 1}^{2N} {{g_r}\phi_z (2x - r)}. \label{eq:s2e4}
\end{equation}
Following this recipe, any scaling sequence $\bm{h}$ that satisfies the orthogonality constraints of Eq. \ref{eq:s2e2} can be applied to generate a set of wavelet functions $\psi_z$ for $z = [1, 2, \ldots]$, which can be argued to form a complete and orthogonal basis for the $1D$ lattice.

In order to realize a wavelet transform that resolves high/low frequencies at each scale additional constraints are imposed on the wavelet sequence $\bm{g} = [g_1, g_2, \ldots, g_{2N}]$, where the specific constraints may depend on the family of wavelets under consideration. For Daubechies wavelets with scaling sequence of $2N$ coefficients, denoted D2N wavelets, it is imposed that the first $N$ moments of the wavelet sequence should vanish, i.e. that
\begin{align}
\sum\limits_{r = 1}^{2N} {\left( {r^\alpha{g_r}} \right)}  = 0, \label{eq:s2e5}
\end{align}
for $\alpha = [0, 1, \ldots, N-1]$.

\subsection{Circuit representation of Daubechies D4 wavelets} \label{sect:D4}
\begin{figure} [!t]
  \begin{centering}
\includegraphics[width=6cm]{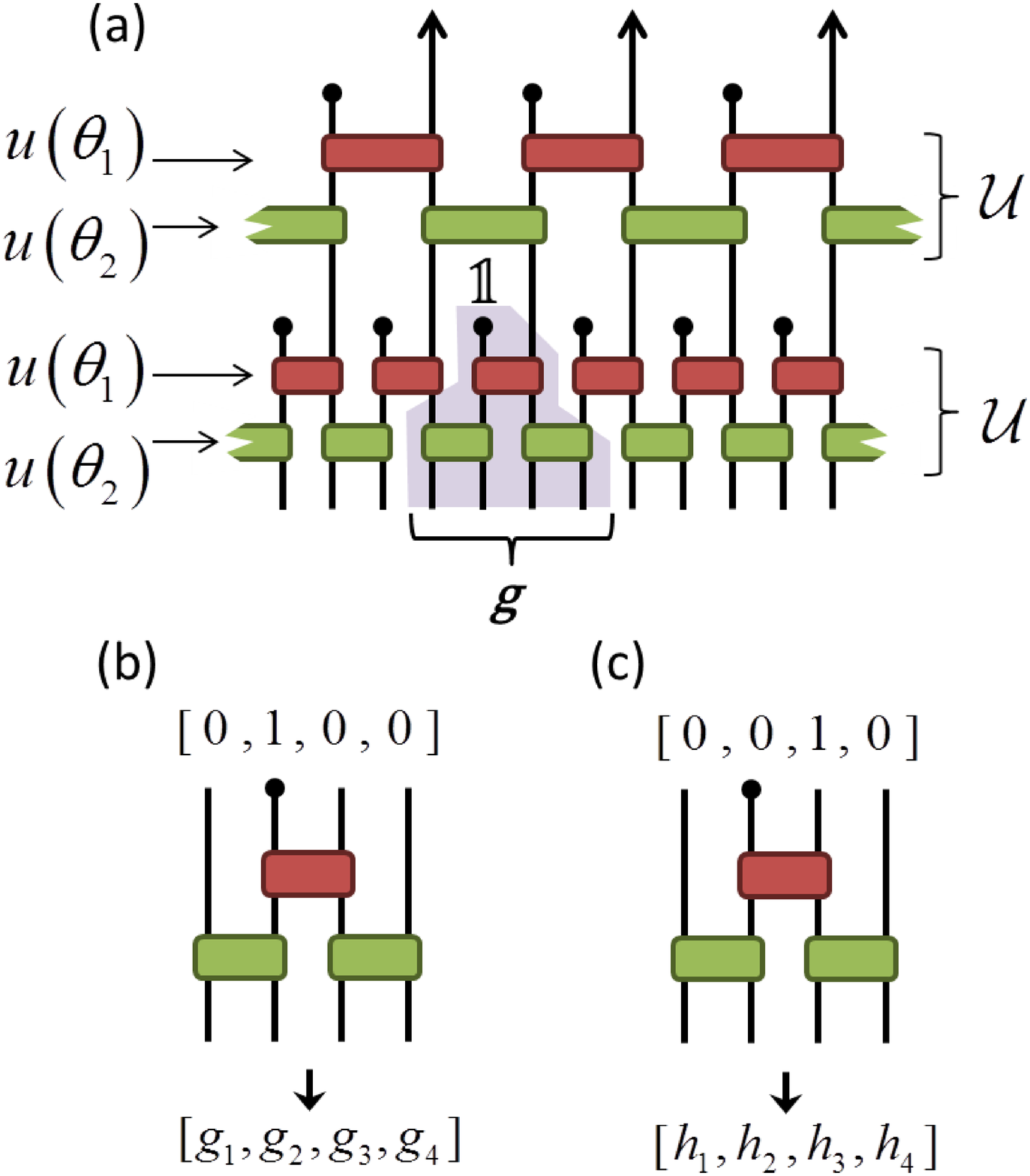}  \end{centering}
  \caption{(a) A multi-scale circuit formed from a binary circuit of depth $N=2$. The wavelet coefficient sequence $\bm{g}$, obtained by transforming the unit vector $\mathbbm 1$, is depicted. (b) The wavelet sequence $\bm{g} = [ g_1, g_2, g_3, g_4 ]$ is obtained by transforming the unit vector ${\mathbbm 1}_r$, with $r$ even (corresponding to an index that terminates), through the binary circuit, see also Eq. \ref{eq:s3e4}. (c) The scaling sequence $\bm{h} = [ h_1, h_2, h_3, h_4 ]$ is obtained by transforming the unit vector ${\mathbbm 1}_r$, with $r$ odd (corresponding to an index that would connect with the next layer of the multi-scale circuit), through the binary circuit, see also Eq. \ref{eq:s3e2}.}
  \label{fig:DepthTwo}
\end{figure}

We now provide an example of how the Daubechies D4 wavelets, which are characterized by the scaling sequence $\bm{h} \approx [ 0.48296, 0.83651, 0.22414, -0.12940]$, can be encoded as a binary unitary circuit $\U$ of depth $N=2$. More precisely, we show that the angles $\{\theta_1, \theta_2 \}$ 
%that parametrize the local unitary matrices $\{u_1, u_2 \}$
that define the circuit $\U$ can be chosen such that the scaling sequence $\bm{h}$ of the D4 wavelets is reproduced when transforming the unit vector $\mathbbm{1}_r$ on any odd site $r$, 
\begin{equation}
\bm{h} = \left\langle \U \times \mathbbm{1}_r \right\rangle,  \label{eq:s3e0}
\end{equation}
where the brackets indicate that we only take the part of the vector with non-zero support, and `$\times$' indicates standard matrix multiplication. Only a few of the local unitary $u$ gates from $\U$ contribute to this non-zero support, see also Fig. \ref{fig:DepthTwo}, such that Eq. \ref{eq:s3e0} may be equivalently expressed as,
%(and similarly the wavelet sequence $\vec g$ is obtained when from the unit vector even sites $r$). That is to say, the scaling sequence is given from the non-zero support of $U \times \mathbbm{1}_r$ for odd $r$, which may be written as,
\begin{equation}
\bm{h} = \left( {{u_2} \oplus {u_2}} \right) \times \left( {1 \oplus {u_1} \oplus 1} \right) \times {[0, 0, 1, 0,]^\dag }, \label{eq:s3e1}
\end{equation}
with `$\oplus$' representing the direct sum. This expression can be evaluated in terms of the tangent angles $t_1$ and $t_2$, see Eq. \ref{eq:s1e3}, that parameterize the local unitary matrices $u(\theta_1)$ and $u(\theta_2) $ respectively, 
\begin{align}
  \left[ {\begin{array}{*{20}{c}}
  {{h_1}} \\ 
  {{h_2}} \\ 
  {{h_3}} \\ 
  {{h_4}} 
\end{array}} \right] &= \left[ {\begin{array}{*{20}{c}}
  1&{{t_2}}&{}&{} \\ 
  { - {t_2}}&1&{}&{} \\ 
  {}&{}&1&{{t_2}} \\ 
  {}&{}&{-{t_2}}&1 
\end{array}} \right]\left[ {\begin{array}{*{20}{c}}
  1&{}&{}&{} \\ 
  {}&1&{{t_1}}&{} \\ 
  {}&{ - {t_1}}&1&{} \\ 
  {}&{}&{}&1 
\end{array}} \right]\left[ {\begin{array}{*{20}{c}}
  0 \\ 
  0 \\ 
  1 \\ 
  0 
\end{array}} \right] \hfill  \nonumber \\
   &= \left[ {\begin{array}{*{20}{c}}
  {{t_1 t_2}} \\ 
  t_1 \\ 
  {  {1}} \\ 
  { -{t_2}} 
\end{array}} \right].\hfill \label{eq:s3e2}
\end{align} 
Note that, for simplicity, we have neglected an overall normalization factor from the $\cos(\theta_k)$ contributions in Eq. \ref{eq:s1e3}. Similarly the wavelet sequence $\bm{g}$, which follows from transforming the unit vector $\mathbbm{1}_r$ on any even site $r$, is given as,
\begin{equation}
\bm{g} = \left( {{u_2} \oplus {u_2}} \right) \times \left( {1 \oplus {u_1} \oplus 1} \right) \times {[0, 1, 0, 0,]^\dag }, \label{eq:s3e3}
\end{equation}
which evaluates to,
\begin{equation}
  \left[ {\begin{array}{*{20}{c}}
  {{g_1}} \\ 
  {{g_2}} \\ 
  {{g_3}} \\ 
  {{g_4}} 
\end{array}} \right] = \left[ {\begin{array}{*{20}{c}}
  {{t_2}} \\ 
  1 \\ 
  { - {t_1}} \\ 
  {  {t_1}{t_2}} 
\end{array}} \right]. \hfill  \label{eq:s3e4}
\end{equation} 
Notice that $\bm{h}$ and $\bm{g}$ satisfy the relation between wavelet and scaling sequences prescribed in Eq. \ref{eq:s2e3} for any choice of the angles $\{\theta_1, \theta_2 \}$. Substituting this parameterization for $\bm{g}$ into Eq. \ref{eq:s2e5} for moments $\alpha = \{0,1\}$ gives,
\begin{align}
  1 - {t_1} + {t_2} + {t_1}{t_2} = 0, \nonumber \\
  2 - 3{t_1} + {t_2} + 4{t_1}{t_2} = 0.  
\end{align}  
These equations are satisfied for the tangent angles,
\begin{equation}
t_1 = 2+\sqrt{3},\phantom{xxx} t_2 = \frac{1}{{\sqrt 3 }}, \label{eq:s3e6}
\end{equation}
%\begin{align}
%  \left[ {{t_1},{t_2}} \right] &= \left[2 + \sqrt 3 , {\tfrac{1}{{\sqrt 3 }}} \right], \label{eq:s3e6}
%\end{align} 
which gives the angles that parameterize the unitary circuit,
\begin{equation}
\theta_1 = \frac{5 \pi }{12},\phantom{xxx} \theta_2 = \frac{{\pi }}{{6}}, \label{eq:s3e5}
\end{equation}
%\begin{equation}
%  \left[ {{\theta _1},{\theta _2}} \right] = \left[ {\tfrac{5 \pi }{12},\tfrac{{\pi }}{{6}}} \right]. \label{eq:s3e5}
%\end{equation}
Substituting these values in Eqs. \ref{eq:s3e2} and \ref{eq:s3e4} for scaling and wavelet sequences (when normalized to unity) yields,
\begin{equation}
\bm{h} \approx \left[ {\begin{array}{*{20}{c}}
 0.48296 \\
 0.83651 \\
 0.22414 \\
-0.12940 
\end{array}} \right],\phantom{xxx} 
  \bm{g} \approx \left[ {\begin{array}{*{20}{c}}
  0.12940 \\ 
  0.22414 \\ 
  -0.83651 \\ 
  0.48296
\end{array}} \right], \label{eq:s3e7} 
\end{equation}
which equates to the known D4 sequences \cite{Daub1}. 

%%%%%%%%%%%%%%%%%%%%%%%%%%%%%%%%%%%%%%%%%%%%%%%%%%%%%%
%%%%%%%%%%%%%%%%%%%%%%%%%%%%%%%%%%%%%%%%%%%%%%%%%%%%%%

\subsection{Circuit representation of generic dyadic wavelets} \label{sect:arbitrary}
We now describe how to construct the unitary circuit representation for generic (compact, orthogonal, dyadic) wavelet transformations. Let $\U$ be a binary unitary circuit of depth $N$, which is parameterized by the set of angles $\{\theta_1, \theta_2,\ldots, \theta_N \}$. We associate scaling $\bm{h}$ and wavelet $\bm{g}$ sequences of $2N$ coefficients to the (non-zero support of the) transformation of the unit vector $\mathbbm{1}_r$ on odd or even sites,
\begin{align}
  \bm{h} &= {\left\langle {\U \times {\mathbbm{1}_r}} \right\rangle},\quad r \; \textrm{odd}  \label{eq:s4e0m} \\
  \bm{g} &= {\left\langle {\U \times {\mathbbm{1}_r}} \right\rangle},\quad r \; \textrm{even}.  \label{eq:s4e0}
\end{align} 
We shall argue the following:
\begin{enumerate}
\item Any scaling $\bm{h}$ and wavelet $\bm{g}$ sequences of $2N$ coefficients (that satisfy the orthogonality constraints of Eq. \ref{eq:s2e2} and are related as in Eq. \ref{eq:s2e3}) can be uniquely encoded as a depth $N$ binary unitary circuit (up to an overall normalization). \label{list:1}
\item Any orthogonal, dyadic WT can be represented as a multi-scale unitary circuit, as described in Eq.\ref{eq:s1e4} and depicted in Fig. \ref{fig:BinCircuit}, which is formed from composition of the appropriate binary unitary circuit that encodes the scaling and wavelet sequences. \label{list:2}
\end{enumerate}
Notice it follows trivially that the scaling $\bm{h}$ coefficient sequences from unitary circuits satisfy the orthogonality constraints of Eq. \ref{eq:s2e2}, as shifts of the scaling sequence correspond to different columns of the unitary matrix $\U$. In Sect. \ref{sect:relate} we offer a proof that $\bm{h}$ and $\bm{g}$ satisfy the relation of Eq. \ref{eq:s2e3}, while in Sect. \ref{sect:algorithm} we describe an algorithm that, for any scaling sequence $\bm{h}$, generates the set of angles $\{\theta_1, \theta_2, \ldots,\theta_N \}$ that parameterizes a depth $N$ unitary circuit that encodes the sequence. Together these two results prove statement \ref{list:1} above. Statement \ref{list:2} then follows easily, as the composition of binary unitary circuits, described in Eq. \ref{eq:s1e4} and depicted in Fig. \ref{fig:BinCircuit}(c), is clearly seen to be equivalent to the refinement equation of Eq. \ref{eq:s2e1}.  

\subsubsection{Relation between scaling and wavelet sequences} \label{sect:relate}
\begin{figure} [!t]
  \begin{centering}
\includegraphics[width=8cm]{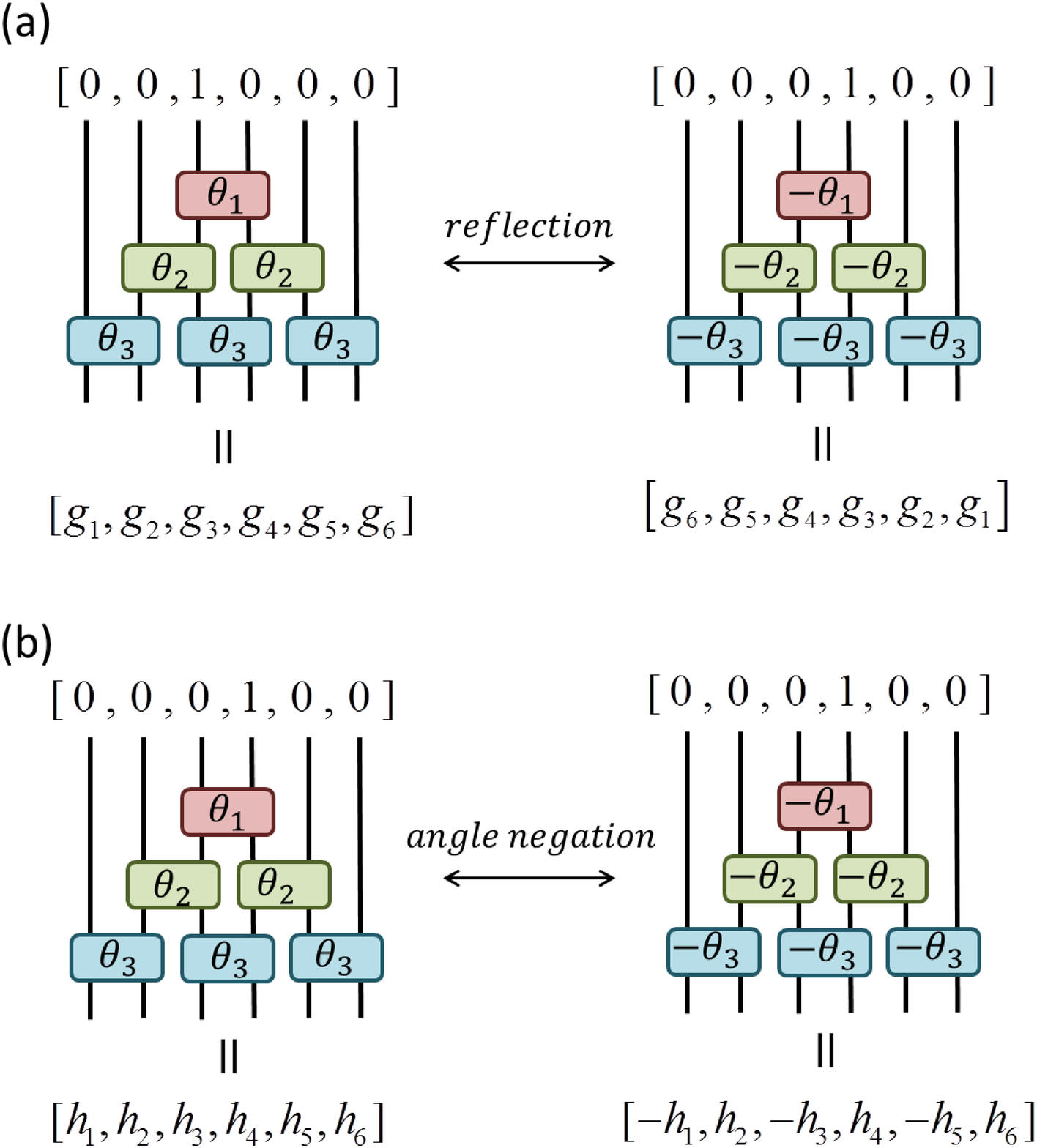}  \end{centering}
  \caption{(a) The wavelet sequence $\bm{g}$ associated to a depth $N=3$ binary circuit. A spatial reflection of the circuit exchanges the positions of the wavelet $\bm{g}$ and $\bm{h}$ scaling sequences, and negates the angles $\theta$ of the local unitary gates. (b) The scaling sequence $\bm{h}$ associated to a depth $N=3$ binary circuit. After negation of the angles $\theta$, the elements of the new scaling sequence $\tilde{\bm{h}}$ relate to the previous as $\tilde{h}_r = \left( { - 1} \right)^{r + 1} h_r$. Together (a) and (b) imply the relation between wavelet $\bm{g}$ and scaling $\bm{h}$ sequences of Eq. \ref{eq:s2e3}.}
  \label{fig:Relation}
\end{figure}

Here we offer a proof that scaling $\bm{h}$ and wavelet $\bm{g}$ sequences from unitary circuits always satisfy the relation of Eq. \ref{eq:s2e3}. We begin by defining the \emph{spatial reflection} operator $\mathcal R$ which reverses the order of both rows and columns of a matrix, i.e if for a given $N\times N$ matrix $A$ we have that $\tilde A \equiv \mathcal{R} \left( A \right)$ then the matrix elements of $A$ and $\tilde A$ are related as
\begin{equation}
{\tilde A}_{i,j} = A_{N-i+1,N-j+1}. \label{eq:s4me1}
\end{equation}
%then the $(i,j)^\textrm{th}$ element of the reflected matrix $\tilde A$ equals the $(N-i+1,N-j+1)^\textrm{th}$ element of $A$.
Notice that spatial reflection of a $2\times 2$ unitary $u$ from Eq. \ref{eq:s1e3} changes the sign of the angle $\theta$,
\begin{equation}
\mathcal{R} \left( {u\left( {{\theta}} \right)} \right) = u\left( { - {\theta}} \right). \label{eq:s4me2}
\end{equation}
It follows that, if $\U$ is a binary unitary circuit parameterized by angles $\{ \theta_1,\theta_2,\ldots,\theta_N \}$, then its spatial reflection $\tilde \U \equiv \mathcal{R} \left( \U \right)$ is a binary unitary circuit with opposite sign angles $\{ -\theta_1,-\theta_2,\ldots,-\theta_N \}$. Notice that spatial reflection of circuit $\U$ into $\tilde \U$ has exchanged the role of the scaling and wavelet sequences, such that if $\bm{g} = [g_1,g_2,\ldots,g_{2N}]$ is the wavelet sequence of $\U$ and $\bm{\tilde h} = [\tilde{h}_1,\tilde{h}_2,\ldots,\tilde{h}_{2N}]$ is the scaling sequence of $\tilde \U$, then
\begin{equation}
{g_r} = {{\tilde h}_{2N - r + 1}}, \label{eq:s4me3}
\end{equation}
see also Fig. \ref{fig:Relation}(a). Thus in order to demonstrate the relation between $\bm{h}$ and $\bm{g}$ of  Eq. \ref{eq:s2e3}, it suffices to show that the elements of $\bm{h}$ and $\tilde{\bm{h}}$ are related with an alternating negative sign,
\begin{equation}
{{\tilde h}_r} = {( - 1)^r}{h_r}, \label{eq:s4me4}
\end{equation}
see also Fig. \ref{fig:Relation}(b), which we now prove. 

Consider the factorization of the circuit into its constituent layers,
\begin{equation}
\U = U_N \times \ldots \times U_2 \times U_1,
\end{equation}
%\begin{equation}
%\U = U(\theta_N) \times \ldots \times U(\theta_2) \times U(\theta_1).
%\end{equation}
where each $U_k$ is a function of some $\theta_k$. Let $\bm{s} = [s_1,s_2,\ldots s_{2M}]$ and $\tilde{\bm{s}} = [\tilde s_1,\tilde s_2,\ldots \tilde s_{2M}]$ be vectors of length $2M$ for integer $M$, and assume ${{\tilde s}_r} = {( - 1)^r}{s_r}$. Define $\bm{s}'$ as given by the transform $\bm{s}$ by circuit layer $U(\theta)$,
\begin{equation}
\bm{s}' \equiv U(\theta ) \times \bm{s} \label{eq:s4me5}
\end{equation}
and $\tilde{\bm{s}}'$ as given by the transform of $\tilde{\bm{s}}$ by a layer $U(-\theta)$ of opposite sign,
\begin{equation}
\tilde{\bm{s}}' \equiv U(-\theta ) \times \tilde{\bm{s}}. \label{eq:s4me6}
\end{equation}
If it can be shown that ${{\tilde s}'_r} = {( - 1)^r}{s'_r}$ then Eq. \ref{eq:s4me4} follows from recursion over all layers $U_k$ of the binary circuit $\U$; in order to show this we evaluate elements of $\bm{s}'$ from Eq. \ref{eq:s4me5},  
\begin{align}
  \left[ {\begin{array}{*{20}{c}}
  {{{s'}_r}} \\ 
  {{{s'}_{r + 1}}} 
\end{array}} \right] &\equiv \left[ {\begin{array}{*{20}{c}}
  {\cos \left( \theta  \right)}&{\sin \left( \theta  \right)} \\ 
  { - \sin \left( \theta  \right)}&{\cos \left( \theta  \right)} 
\end{array}} \right]\left[ {\begin{array}{*{20}{c}}
  {{s_r}} \\ 
  {{s_{r + 1}}} 
\end{array}} \right] \hfill \nonumber \\
   & = \left[ {\begin{array}{*{20}{c}}
  {{s_r}\cos \left( \theta  \right) + {s_{r + 1}}\sin \left( \theta  \right)} \\ 
  { - {s_r}\sin \left( \theta  \right) + {s_{r + 1}}\cos \left( \theta  \right)} 
\end{array}} \right], \label{eq:s4me7}
\end{align} 
and similarly evaluate elements $\tilde{\bm{s}}'$ from Eq. \ref{eq:s4me6}, 
\begin{align}
  \left[ {\begin{array}{*{20}{c}}
  {{{\tilde s'}_r}} \\ 
  {{{\tilde s'}_{r + 1}}} 
\end{array}} \right] & \equiv \left[ {\begin{array}{*{20}{c}}
  {\cos \left( \theta  \right)}&{ - \sin \left( \theta  \right)} \\ 
  {\sin \left( \theta  \right)}&{\cos \left( \theta  \right)} 
\end{array}} \right]\left[ {\begin{array}{*{20}{c}}
  {{{\tilde s}_r}} \\ 
  {{{\tilde s}_{r + 1}}} 
\end{array}} \right] \hfill \nonumber \\
   & = \left[ {\begin{array}{*{20}{c}}
  {{{\tilde s}_r}\cos \left( \theta  \right) - {{\tilde s}_{r + 1}}\sin \left( \theta  \right)} \\ 
  {{{\tilde s}_r}\sin \left( \theta  \right) + {{\tilde s}_{r + 1}}\cos \left( \theta  \right)} 
\end{array}} \right] \hfill \nonumber \\
   & = \left[ {\begin{array}{*{20}{c}}
  { - {s_r}\cos \left( \theta  \right) - {s_{r + 1}}\sin \left( \theta  \right)} \\ 
  { - {s_r}\sin \left( \theta  \right) + {s_{r + 1}}\cos \left( \theta  \right)} 
\end{array}} \right]\nonumber \\ 
& = \left[ {\begin{array}{*{20}{c}}
  { - {{s'}_r}} \\ 
  {{{s'}_{r + 1}}} 
\end{array}} \right], \label{eq:s4me8}
\end{align}
from which the desired result is observed. Thus we conclude that the scaling $\bm{h}$ and wavelet $\bm{g}$ sequences from a binary unitary circuit are always related as per Eq. \ref{eq:s2e3}.

%%%%%%%%%%%%%%%%%%%%%%%%%%%%%%%%%%%%%%%%%%%%%%%%%%%%%%
%%%%%%%%%%%%%%%%%%%%%%%%%%%%%%%%%%%%%%%%%%%%%%%%%%%%%%

\subsubsection{Circuit construction algorithm} \label{sect:algorithm}
Given a discrete, orthogonal WT described by the scaling sequence $\bm{h} = [h_1, h_2, \ldots, h_{2N}]$ of $2N$ coefficients, we now describe an algorithm to encode the sequence as a depth $N$ binary unitary circuit. More precisely, we describe how the angles $\{\theta_1, \theta_2,\ldots, \theta_N \}$ defining the circuit can be uniquely chosen such that the scaling sequence is mapped to a unit vector $\mathbbm{1}_r$ under transformation by the circuit, 
\begin{equation}
\bm{h}^{\dag} \times \U = (\mathbbm{1}_r)^{\dag}, \label{eq:s4e1}
\end{equation}
in accordance with the definition of Eq. \ref{eq:s4e0m}. Recall that the binary circuit decomposes into $N$ layers, $\U = U_N U_{N-1} \ldots U_1$, where each layer $U_k$ is the direct sum of a $2\times 2$ unitary matrices $u(\theta_k)$. Let ${\bm{h}}'$ be the scaling sequence after transforming by the bottom layer $U_N$ of the unitary circuit,
\begin{equation}
({\bm{h}}')^{\dag} \equiv \bm{h}^{\dag} \times U_N , \label{eq:s4e2}
\end{equation}
We now propose that the free angle $\theta_N$ of unitary $u(\theta_N)$ should be chosen as, 
\begin{equation}
{\theta} = \left\{ {\begin{array}{*{20}{l}}
  \textrm{(i)} &{{{\arctan }}\left( {\frac{h_1}{h_2}} \right)}, &{{h_2} \ne 0} \\ 
  \textrm{(ii)} &{{{\arctan }}\left( \frac{-h_{2N}}{h_{2N - 1}} \right)}, &{{h_2} = 0,\; {h_{2N - 1}} \ne 0} \\ 
  \textrm{(iii)}\; &{\pi /2}, &{{h_2} = {h_{2N - 1}} = 0.} 
\end{array}} \right. \label{eq:s4e3}
\end{equation}
Under this choice the first element of the transformed scaling sequence ${\bm{h}}'$ is zero,
\begin{equation}
h_1' = h_1 \cos(\theta) - h_2 \sin(\theta) = 0. \label{eq:s4e4}
\end{equation}
for all three possibilities (i-iii) from Eq. \ref{eq:s4e3}. To understand this in the second instance (ii) notice that the orthogonality constraints of Eq. \ref{eq:s2e2} impose,
\begin{equation}
h_1 h_{2N-1} + h_2 h_{2N} = 0, \label{eq:s4e5}
\end{equation}
which implies that $h_1=0$ given that $h_{2N-1} \ne 0$. Similarly, the trailing $(2N)^\textrm{th}$ element of ${\bm{h}}'$ is also mapped to zero,
\begin{equation}
h_{2N}' =  h_{2N-1} \sin(\theta) + h_{2N} \cos(\theta) = 0. \label{eq:s4e6}
\end{equation}
In instances (ii) and (iii) from Eq. \ref{eq:s4e3} this follows trivially, whereas in the first instance (i) this can be understood by substituting the orthogonality constraint of Eq. \ref{eq:s2e2} to give,
\begin{align}
h_{2N}' & = h_{2N} \left( -h_2 / h_1  \sin(\theta) + \cos(\theta)\right),\nonumber \\
        & = \left( h_{2N} h_1' \right) / h_1 = 0. \label{eq:s4e7}
\end{align}
Thus we have demonstrated that under the choice of angle $\theta$ from Eq. \ref{eq:s4e3} the scaling sequence $\bm{h}$ of $2N$ coefficients is mapped under $U_N$ to a new sequence $\bm{h} '$ of $2N - 2$ coefficients (where the first and last elements of $\bm{h} '$, which were shown to be zero, have been dropped). It can also easily be seen that the orthogonality constraints on $\bm{h}$, see Eq. \ref{eq:s2e2}, map to an equivalent set of constraints for the new sequence $\bm{h} '$. The set of angles $\{\theta_N, \theta_{N-1}, \ldots, \theta_2, \theta_1 \}$ is obtained by iterating this procedure a total of $N-1$ times, i.e. for unitary layers $\{U_{N}, U_{N-1}, \ldots, U_2\}$, which then maps the initial scaling sequence $\bm{h}$ of $2N$ coefficients into a sequence of length two denoted $\bm{h}^\textrm{f} = [h_1^\textrm{f}, h_2^\textrm{f}]$. At this stage, the angle $\theta_1$ associated to the top level $U_1$ is fixed at,
\begin{equation}
\theta_1 = \tan^{-1} \left( h_1^\textrm{f} / h_2^\textrm{f} \right), \label{eq:s4e8}
\end{equation}
such that $\bm{h}^\textrm{f}$ is mapped to a sequence with only a single non-zero element, indicating that the scaling sequence is encoded in the binary unitary circuit.

\begin{figure} [!t]
  \begin{centering}
\includegraphics[width=6cm]{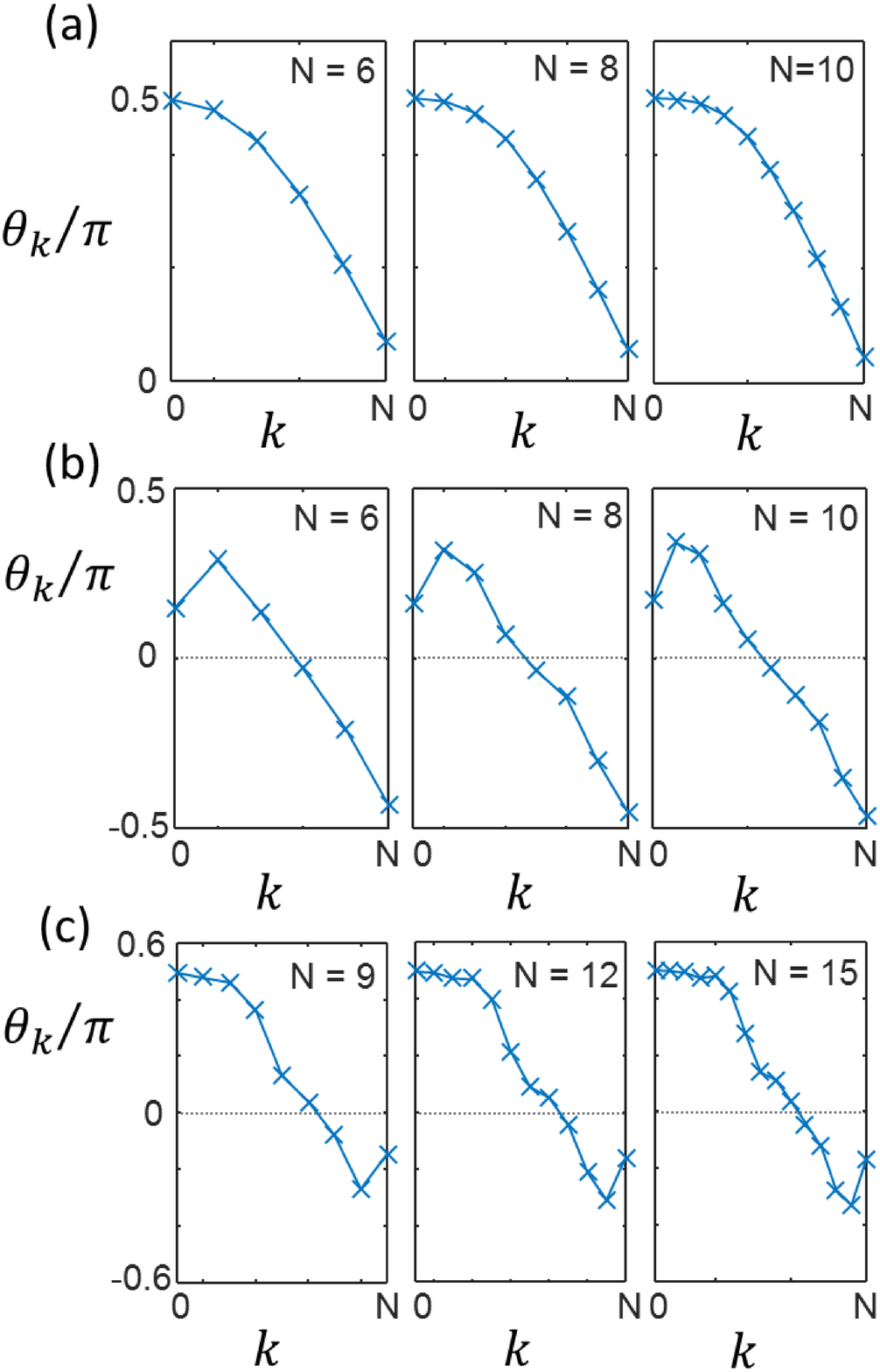}  \end{centering}
  \caption{Plots of the angles $\theta_k$ that parameterize the circuit representations of (a) the D2N Daubechies wavelets for $N = \{6,8,10\}$, (b) the symlets of $2N$ coefficients for $N = \{6,8,10\}$, and (c) the coiflets of $2N$ coefficients for $N = \{9,12,15\}$.} \label{fig:Angles}
\end{figure}

Employing this algorithm we generate the sets of angles that describe the Daubechies wavelets D2N for $N=\{1,2,\ldots 10\}$, see Tab. \ref{Tab:Daubechies}, the symlets of $2N$ coefficients for $N=\{1,2,\ldots 10\}$, see Tab. \ref{Tab:Symlets}, and the coiflets of $2N$ coefficients for $N=\{3,6,9,12,15\}$, see Tab. \ref{Tab:Coiflets}. Notice that clear patterns are evident in the angles $\theta$ between different orders $N$ of wavelets in each of the families, see also Fig. \ref{fig:Angles}.

%%%%%%%%%%%%%%%%%%%%%%%%%%%%%%%%%%%%%%%%%%%%%%%%%%%%%%
%%%%%%%%%%%%%%%%%%%%%%%%%%%%%%%%%%%%%%%%%%%%%%%%%%%%%%

\begin{table*}[!p]
\centering
\begin{tabular}{|c|c|c|c|c|c|c|c|c|c|c|}
\hline
    & $\;$ D2 $\;$& $\;$ D4 $\;$&\; D6 $\;$ &$\;$D8 $\;$ &$\;$D10 $\;$ &$\;$ D12 $\;$  &$\;$D14 $\;$ &$\;$ D16 $\;$ & $\;$D18   $\;$ &$\;$ D20 $\;$ \\ \hline
$\theta_1/\pi$  & $\;$1/4$\;$ & $\;$5/12$\;$ & $\;$0.466419$\;$ & $\;$0.485368$\;$ & $\;$0.493368$\;$ &$\;$ 0.496925 $\;$& $\;$0.498553 $\;$&$\;$ 0.499312$\;$ & $\;$0.499671$\;$ & $\;$ 0.499841 $\;$ \\
$\theta_2/\pi$  &  -   & 1/6  & 0.340895 & 0.419242 & 0.457854 & 0.477754 & 0.488217 & 0.493760 & 0.496701 & 0.498260 \\
$\theta_3/\pi$  &  -   &  -    & 0.124476 & 0.283112 & 0.372126 & 0.424117 & 0.455055 & 0.473548& 0.484560 & 0.491068 \\
$\theta_4/\pi$  &  -   &   -   &    -      & 0.099238 & 0.240141 & 0.330420 & 0.389528 & 0.428559 & 0.454286 & 0.471103 \\
$\theta_5/\pi$  &  -   &   -   &    -      &     -              & 0.082499 & 0.207731 & 0.295133 & 0.357027 & 0.401092 & 0.432370 \\
$\theta_6/\pi$  &  -   &   -   &    -      &     -              &      -             & 0.070599 & 0.182705 & 0.265668 & 0.327768 & 0.374473 \\
$\theta_7/\pi$  &  -   &   -   &    -      &     -              &      -             &     -              & 0.061708 & 0.162913 & 0.241041 & 0.301941 \\
$\theta_8/\pi$  &  -   &   -   &    -      &     -              &      -             &     -              &                  - & 0.054813 & 0.146919 & 0.220319 \\
$\theta_9/\pi$  &  -   &   -   &    -      &     -              &      -             &     -              &                  - &          -         & 0.049309 & 0.133750 \\
$\theta_{10}/\pi$ &  -   &   -   &    -      &     -              &      -             &     -              &                  - &         -          &        -           & 0.044815 \\ \hline
\end{tabular}
\caption{Set of angles $\theta$ that parameterize the depth $N$ binary unitary circuit corresponding to the D2N Daubechies wavelets.}
\label{Tab:Daubechies}
\end{table*}

\begin{table*}[!p]
\centering
\begin{tabular}{|c|c|c|c|c|c|c|c|c|c|c|}
\hline
    & $N=1$  & $N=2$   & $N=3$       & $N=4$       & $N=5$     & $N=6$   & $N=7$  & $N=8$   & $N=9$  & $N=10$     \\ \hline
$\theta_1/\pi$  & $\;$1/4 $\;$& $\;$5/12$\;$ & $\;$0.466419$\;$ &$\;$ 0.128000  $\;$& $\;$0.197549  $\;$& $\;$0.149210  $\;$& $\;$0.418681 $\;$ & $\;$0.162192 $\;$ & $\;$0.207549 $\;$ & $\;$0.171178 $\;$ \\
$\theta_2/\pi$  &  -   & 1/6  & 0.340895 & 0.213974  & -0.086984 & 0.289866  & -0.195828 & 0.323040  & -0.481080 & 0.343162 \\
$\theta_3/\pi$  &  -   &  -    & 0.124476 & -0.045343 & 0.279718  & 0.138289  & 0.068910 & 0.252744  & -0.018742 & 0.308034  \\
$\theta_4/\pi$  &  -   &  -    &    -      & -0.381317 & 0.076487  & -0.028270 & 0.498795  & 0.071840  & 0.380963  & 0.163525  \\
$\theta_5/\pi$  &  -   &  -    &    -      &         -           & -0.237764 & -0.204970 & -0.277830 & -0.032859 & 0.331334 & 0.059282  \\
$\theta_6/\pi$  &  -   &  -    &    -      &         -           &          -          & -0.429066 & 0.038275  & -0.112982 & 0.180695  & -0.027548 \\
$\theta_7/\pi$  &  -   &  -    &    -      &         -           &          -          &         -           & 0.381482  & -0.299591 & 0.086208  & -0.104053 \\
$\theta_8/\pi$  &  -   &  -    &    -      &         -           &          -          &         -           &         -           & -0.449411 & -0.091644 & -0.185998 \\
$\theta_9/\pi$  &  -   &  -    &    -      &         -           &          -          &         -           &         -           &           -         & -0.367415 & -0.351975 \\
$\theta_{10}/\pi$ & -    & -     &  -        &       -             &        -            &       -             &       -             &         -           &      -              & -0.460675 \\ \hline
\end{tabular}
\caption{Set of angles $\theta$ that parameterize the depth $N$ binary unitary circuit corresponding to the symlets of $2N$ coefficients.}
\label{Tab:Symlets}
\end{table*}

\begin{table*}[!p]
\centering
\begin{tabular}{|c|c|c|c|c|c|c|c|c|c|c|}
\hline
    & $N=3$                 & $N=6$                & $N=9$                & $N=12$ & $N=15$                \\ \hline
$\theta_1/\pi$   &$\;$ 0.432513$\;$  &$\;$ 0.486012$\;$  &$\;$ 0.497096$\;$  & $\;$ 0.499363 $\;$ & $\;$ 0.499855 $\;$  \\
$\theta_2/\pi$   & 0.115026  & 0.449082  & 0.478861  & 0.494102  & 0.498385  \\
$\theta_3/\pi$   & -0.067486 & 0.293308  & 0.462006  & 0.474920  & 0.491529  \\
$\theta_4/\pi$   &         -           & 0.036930  & 0.365035  & 0.471844  & 0.472624  \\
$\theta_5/\pi$   &         -           & -0.163110 & 0.132236  & 0.401463  & 0.479637  \\
$\theta_6/\pi$   &         -           & -0.119802 & 0.040847  & 0.216145  & 0.423857  \\
$\theta_7/\pi$   &         -           &         -           & -0.080093 & 0.091539  & 0.274343  \\
$\theta_8/\pi$   &         -           &         -           & -0.267867 & 0.053920  & 0.139303  \\
$\theta_9/\pi$   &         -           &         -           & -0.144368 & -0.041439 & 0.111993  \\
$\theta_{10}/\pi$  &       -             &       -             &      -              & -0.209282 & 0.038183  \\
$\theta_{11}/\pi$  &       -             &       -             &      -              & -0.308590 & -0.041666 \\
$\theta_{12}/\pi$  &       -             &       -             &      -              & -0.159473 & -0.122667 \\
$\theta_{13}/\pi$  &       -             &       -             &      -              &           -         & -0.278205 \\
$\theta_{14}/\pi$  &       -             &       -             &      -              &           -         & -0.332208 \\
$\theta_{15}/\pi$  &       -             &       -             &      -              &           -         & -0.170007 \\ \hline
\end{tabular}
\caption{Set of angles $\theta$ that parameterize the depth $N$ binary unitary circuit corresponding to the coiflets of $2N$ coefficients.}
\label{Tab:Coiflets}
\end{table*}

\section{Circuit constructions of novel wavelets} \label{sect:novel}
In the remainder of this manuscript we discuss the use of generalized unitary circuits for the design and realization of novel wavelet families. This includes families of (dilation factor $m = 3$) symmetric wavelets in Sect. \ref{sect:dilation3}, symmetric multiwavelets in Sect. \ref{sect:multi}, a family of (dilation factor $m = 4$) symmetric wavelets in Sect. \ref{sect:dilation4}, boundary wavelets in Sect. \ref{sect:boundary}, and symmetric biorthogonal wavelets in Sect. \ref{sect:biorthog}. Many of the generalized circuits considered are direct analogues of circuits considered previously in the context of MERA for quantum many-body systems (for instance, of the reflection symmetric MERA considered in Refs. \onlinecite{MERAapp1,MERAapp2}, of the MERA of different dilation factors considered in Refs. \onlinecite{MERAbook,Alg}, and of the boundary MERA considered in Refs. \onlinecite{BoundMERA1,BoundMERA2}).

\subsection{Dilation $m=3$ symmetric wavelets} \label{sect:dilation3}
An often desirable characteristic of wavelets is symmetry (or antisymmetry) under spatial reflections. However it is known that the only compactly supported orthogonal wavelet basis of dilation factor $m = 2$ that consists of symmetric (or antisymmetric) functions is the trivial Haar basis\cite{Daub1,Daub2}. One solution to this problem is to use a larger dilation factor $m > 2$, which results in a scheme with one scaling and $(m-1)$ wavelet functions. Constructions of wavelets with larger dilation factors have been studied extensively in previous works, see for instance Refs. \onlinecite{Sym1,Sym2,Sym3,Sym4}. In this Section we consider the design of novel symmetric (and/or antisymmetric) wavelets with dilation factor $m = 3$.

In the context of MERA quantum circuits, reflection symmetry (and, more generally, other spatial symmetries as well as global internal symmetries), are imposed by requiring that the individual unitary gates that comprise the circuit satisfy the symmetry\cite{MERAapp1,MERAapp2}. Here we follow a similar strategy, and construct reflection symmetric wavelets using unitary circuits where all of the unitary matrices that comprise the circuit are individually symmetric under spatial reflections. For $2\times 2$ unitary matrices, there is only a single non-trivial instance of a matrix, which we denote $u_\textrm{sw}$, that is invariant under reflections $\mathcal R$ (as defined in Eq. \ref{eq:s4me1}), 
\begin{equation}
u_\textrm{sw} = \left[ {\begin{array}{*{20}{c}}
  0&1 \\ 
  1&0 
\end{array}} \right], \label{eq:s5e2}
\end{equation}
which simply enacts a swap of elements on a length $2$ vector. For $3\times 3$ unitary matrices, a 1-parameter family of reflection symmetric matrices $v(\theta)$ is obtained by exponentiating the most general instance of a real skew-hermitian matrix that is also reflection symmetric,
\begin{equation}
  v\left( \theta  \right) = \exp \left( {\frac{1}{{\sqrt 2 }}\left[ {\begin{array}{*{20}{c}}
  0&\theta &0 \\ 
  { - \theta }&0&{ - \theta } \\ 
  0&\theta &0 
\end{array}} \right]} \right), \label{eq:s5e3}
\end{equation}
which evaluates to,
\begin{equation}
 v\left( \theta  \right) = \frac{1}{2}\left[ {\begin{array}{*{20}{c}}
  {\cos \left( \theta  \right) + 1}&{\sqrt 2 \sin \left( \theta  \right)}&{\cos \left( \theta  \right) - 1} \\ 
  { - \sqrt 2 \sin \left( \theta  \right)}&{2\cos \left( \theta \right)}&{ - \sqrt 2 \sin \left( \theta  \right)} \\ 
  {\cos \left( \theta  \right) - 1}&{\sqrt 2 \sin \left( \theta  \right)}&{\cos \left( \theta  \right) + 1} 
\end{array}} \right] \hfill. \label{eq:s5e4}
\end{equation} 
We now describe how to build a unitary circuit from these $3\times 3$ reflection symmetric matrices $v\left( \theta  \right)$ in conjunction with the swap gate $u_\textrm{sw}$, that will subsequently be used to parameterize a family of symmetric wavelet transforms with dilation factor $m=3$.

%%%%%%%%%%%%%%%%%%%%%%%%%%%%%%%%%%%%%%%%%%%%%%%%%%%%%%
\begin{figure} [!h]
  \begin{centering}
\includegraphics[width=8.5cm]{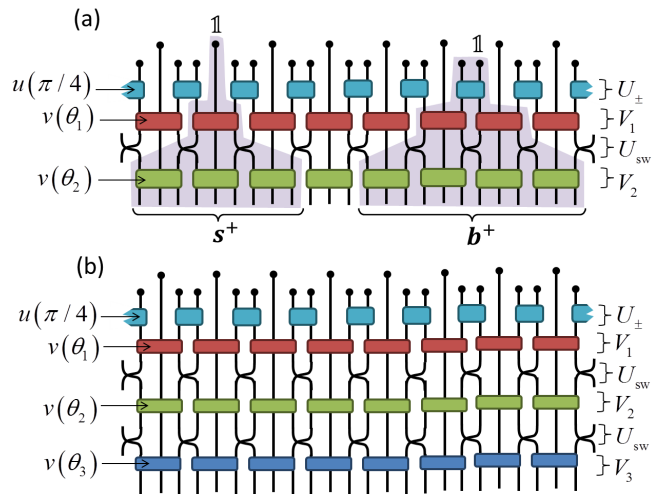}  \end{centering}
  \caption{(a) A depth $N = 2$ ternary unitary circuit. Layers $V_k$ are composed of a direct sum of $3\times 3$ reflection symmetric matrices $v$ of Eq.\ref{eq:s5e4}, which are intersperced with layers $U_\textrm{sw}$ composed of two site swap gates $u_\textrm{sw}$ of Eq.\ref{eq:s5e2} (depicted as the crossing of wires). The top layer of the circuit $U_{\pm}$ is composed of $2\times 2$ matrices $u$ of Eq.\ref{eq:s1e3} with angle $\theta = \pi/4$, which creates reflection symmetric / antisymmetric outputs from input unit vectors. Tranforming a unit vector $\mathbbm 1$ located at the central site of a $v$ matrix gives (site-centered) symmetric sequence $\bm{s}^+$, while a unit vector on the right site of a $u$ matrix gives (edge-centered) symmetric sequence $\bm{b}^+$ and a unit vector on the left site of a $u$ matrix gives (edge-centered) antisymmetric sequence $\bm{b}^-$ (not shown). (b) A depth $N = 3$ ternary unitary circuit.} \label{fig:TernCircuit}
\end{figure}
%%%%%%%%%%%%%%%%%%%%%%%%%%%%%%%%%%%%%%%%%%%%%%%%%%%%%%

\subsubsection{Ternary unitary circuits} 
Let $\V$ be a $3M \times 3M$ matrix for integer $M$; we say that $\V$ is a \emph{ternary unitary circuit} of depth $N$ if it can be decomposed as, 
\begin{equation}
\V = {V_N} U_\textrm{sw} {V_{N - 1}} U_\textrm{sw} \ldots {V_2} U_\textrm{sw} {V_1} U_\pm,
\end{equation}
see also the circuit diagram of Fig. \ref{fig:TernCircuit}. Here each of the $N$ sublayers $V_k$ is a direct sum of $M$ matrices $v(\theta_k)$ from Eq. \ref{eq:s5e4},
\begin{equation}
V_k = \ldots \oplus v(\theta_k) \oplus v(\theta_k) \oplus v(\theta_k) \oplus \ldots, \label{eq:s5e5}
\end{equation}
which are interspersed with sublayers $U_\textrm{sw}$ consisting of direct sums of swap gates $u_\textrm{sw}$ from Eq. \ref{eq:s5e2},
\begin{equation}
U_\textrm{sw} = \ldots \oplus u_\textrm{sw} \oplus \mathbb{I} \oplus u_\textrm{sw} \oplus \mathbb{I} \oplus u_\textrm{sw} \oplus \mathbb{I} \oplus \ldots, \label{eq:s5e7}
\end{equation}
with $\mathbb{I}$ the $1\times 1$ identity. The ternary circuit $\V$ includes a top sublayer $U_{\pm}$ composed of $2\times 2$ unitary gates $u(\pi/4)$ which are spaced by contributions of the identity, 
\begin{equation}
U_{\pm} = \ldots \oplus u(\pi/4) \oplus \mathbb{I} \oplus u(\pi/4) \oplus \mathbb{I} \oplus \ldots. \label{eq:s5e8}
\end{equation}

\begin{figure} [!h]
  \begin{centering}
\includegraphics[width=8cm]{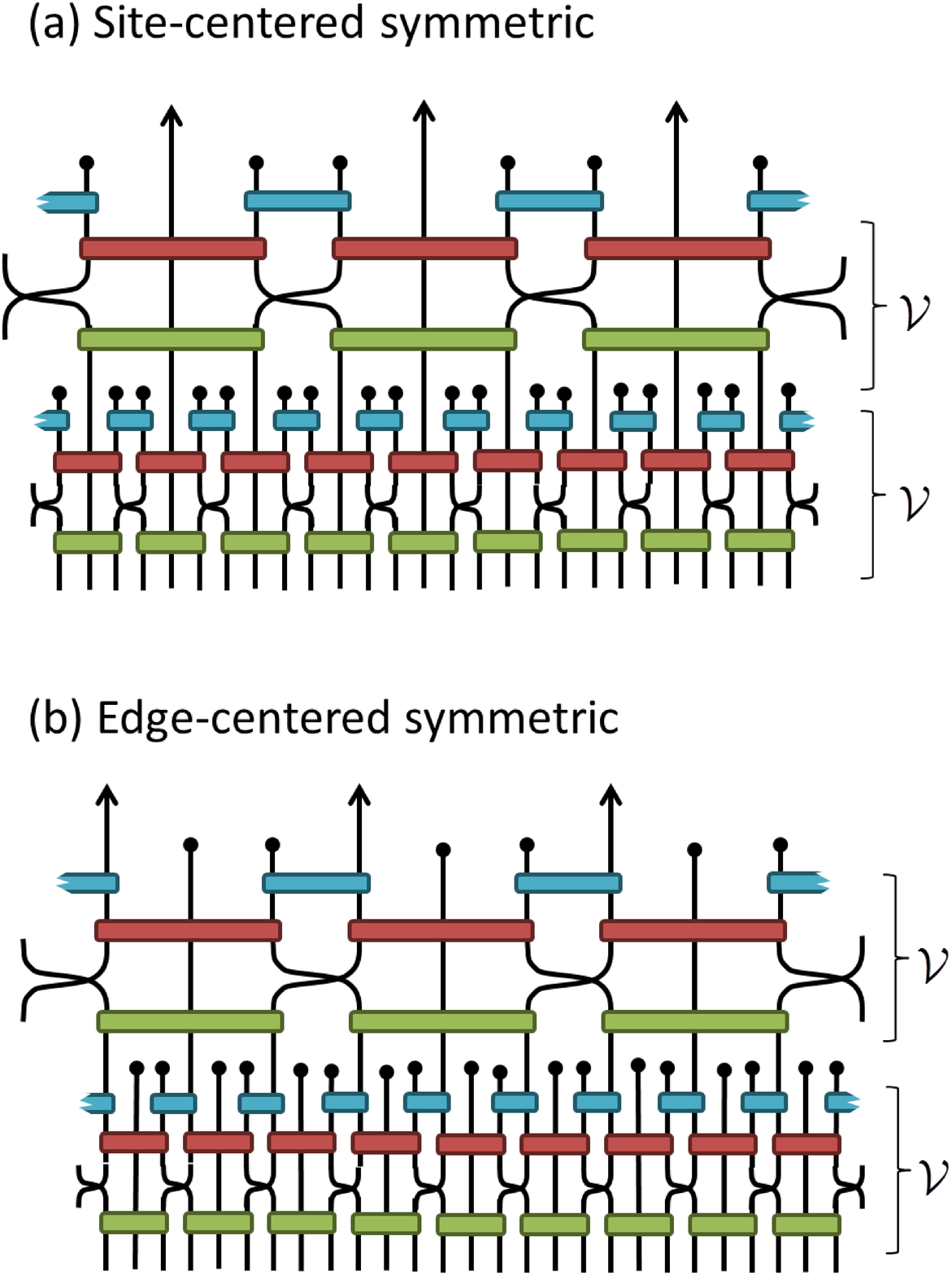}  \end{centering}
  \caption{Two different types of multi-scale circuit formed from composition of depth $N=2$ ternary unitary circuits. (a) The site-centered multi-scale circuit is given when treating sequence $\bm{s}^+$ of Fig. \ref{fig:TernCircuit} as the scaling function. (b) The edge-centered multi-scale circuit is given when treating sequence $\bm{b}^+$ of Fig. \ref{fig:TernCircuit} as the scaling function. Both types of multiscale circuit describe orthogonal, dilation factor 3, discrete wavelet transforms with exactly symmetric / antisymmetric wavelets.} \label{fig:MultiTern}
\end{figure}

Notice that the ternary unitary circuit of depth $N$ is fully parameterized by the set of angles $\{\theta_1, \theta_2, \ldots, \theta_N \}$. 
%The ternary circuit is constructed such that, for any choice of angles $\theta_k$, the associated scaling and wavelet sequences either exactly symmetric or antisymmetric w.r.t spatial reflections.
There are three coefficient sequences associated to a ternary circuit (which corresponds to a three-channel filter bank), which we label $\{ \bm{s}^ +,\bm{b}^ +,\bm{b}^- \}$. Each of the three sequences is obtained by transforming a unit vector $\mathbbm{1}_r$ on different sites $r$ with the ternary circuit $\V$,
\begin{equation}
\left\langle {\V \times {\mathbbm{1}_r}} \right\rangle = \left\{ {\begin{array}{*{20}{l}}
  {\textrm{(i)}}&{{\bm{s}^ + },}\; \;   &{r = 2,5,8, \ldots } \\ 
  {\textrm{(ii)}}&{{\bm{b}^ + },}\; \;  &{r = 3,6,9, \ldots } \\ 
  {\textrm{(iii)}}&{{\bm{b}^ - },}\; \; &{r = 1,4,7, \ldots } 
\end{array}} \right., \label{eq:s5e10}
\end{equation}
see also Fig. \ref{fig:TernCircuit}. The sequences possess the following properties:
\begin{itemize}
\item $\bm{s}^ +$ is a site-centered symmetric sequence of (odd) length $6N-3$ elements (i.e. that is symmetric w.r.t. reflections about its central element)
\item $\bm{b}^ +$ is a edge-centered symmetric sequence of (even) length $6N$ elements (i.e. that is symmetric w.r.t. reflections centered between its two central elements)
\item $\bm{b}^ -$ is a edge-centered antisymmetric sequence of (even) length $6N$ elements (i.e. that is antisymmetric w.r.t. reflections centered between its two central elements).
\end{itemize}
Similar to the construction applied to the binary unitary circuit in Sect. \ref{sect:circuit}, layers of the ternary circuit can be composed to form a multi-scale circuit,
\begin{equation} 
\V_1 \circ \V_2 \circ \V_3 \circ \V_4 \circ \ldots, \label{eq:s5e11}
\end{equation} 
which encodes the dilation factor 3 discrete WT. Notice that there are two distinct ways of implementing this composition dependent on whether $\bm{s}^+ $ or $\bm{b}^+$ is considered as the scaling sequence, which we refer to as the site-centered or edge-centered multi-scale circuits respectively, see also Fig. \ref{fig:MultiTern}. 

%%%%%%%%%%%%%%%%%%%%%%%%%%%%%%%%%%%%%%%%%%%%%%%%%%%%%%
%%%%%%%%%%%%%%%%%%%%%%%%%%%%%%%%%%%%%%%%%%%%%%%%%%%%%%

\subsubsection{Example 1: dilation $m=3$ wavelets with maximal vanishing moments} \label{sect:ex1}
A depth $N$ ternary unitary circuit offers a parameterization of exactly reflection symmetric and antisymmetric wavelets with $N$ free parameters $\{\theta_1, \theta_2, \ldots, \theta_N \}$. In this section we consider the family of wavelets that result from choosing these parameters to maximize the number of vanishing moments that the wavelets possess. 

We use the site-centered form of the multi-scale circuit as depicted in Fig. \ref{fig:MultiTern}(a), where sequence $\bm{s}^+$ is treated as the scaling sequence and $\bm{b}^{+}$, $\bm{b}^{-}$ are the wavelet sequences, and we optimize the angles $\theta_k$ such the wavelets have vanishing moments,
\begin{equation}
\sum\limits_{r} {\left( {r^\alpha {b_r^{+}}} \right)}  = 0,\phantom{xxx} \sum\limits_{r} {\left( {r^\alpha {b_r^{-}}} \right)}  = 0,  \label{eq:s5e12}
\end{equation}
for $\alpha=[0,1,2,\ldots ]$. Notice that $\bm{b}^{+}$ is automatically orthogonal to odd $\alpha$ polynomials due to its symmetry, and similarly $\bm{b}^{-}$ is automatically orthogonal to even $\alpha$ polynomials. Thus satisfying Eq. \ref{eq:s5e12} for orders $\alpha=[0,1,2,\ldots,N-1]$, i.e. such that the wavelets have $N$ vanishing moments, requires $N$ non-trivial constraints to be met.

The sets of angles $\{\theta_1, \theta_2, \ldots, \theta_N \}$ that satisfy the vanishing moment criteria for circuit depths $N = 2,4,6$ are given in Tab. \ref{table:MaxTernary}. These were obtained numerically using the standard Nelder-Mead algorithm to minimize the cost function given by the sum of the first $N$ moments of the wavelet sequences. The resulting wavelets are depicted in Fig. \ref{fig:MaxTern}. Interestingly, while the scaling functions associated to these WTs appear similar to those obtained previously in Figs. 3 and 5 of Ref. \onlinecite{Sym1}, the wavelets themselves are seen to have much smaller effective support (i.e. the region over which the wavelets have non-vanishingly small amplitude is reduced), which suggests they may perform better in practical applications.

%%%%%%%%%%%%%%%%%%%%%%%%%%%%%%%%%%%%%%%%%%%%%%%%%%%%%%%%%%%%%%%%%%%%%%%%%%%%%%%
\begin{table}[!htb]
\begin{tabular}{|c||c|c|c|}
\hline
                
  & \phantom{xxx} $N=2$ \phantom{xxx}&\phantom{xxx} $N=4$\phantom{xxx} &\phantom{xxx} $N=6$ \phantom{xxx}\\ \hline
	
$\theta_1 $ & 0.275642799  &  0.595157579  & 0.756972477 \\ 	
$\theta_2 $ & 0.679673818  & -0.840085482  &-1.401172929 \\ 
$\theta_3 $ & -            & -0.314805259  & 0.537202982 \\
$\theta_4 $ & -            &  1.515049781  & 0.264416395 \\	
$\theta_5 $ & -            & -             &-1.007834534 \\	
$\theta_6 $ & -            & -             & 1.805732227 \\										
\hline
\end{tabular}
\caption{Angles $\theta_k$ parameterizing the site-centered ternary circuits of depths $N = \{2,4,6 \}$ such that the wavelets have $N$ vanishing moments.}
\label{table:MaxTernary}
\end{table}
%%%%%%%%%%%%%%%%%%%%%%%%%%%%%%%%%%%%%%%%%%%%%%%%%%%%%%%%%%%%%%%%%%%%%%%%%%%%%%%

\begin{table}[!h!tb]
\centering
\begin{tabular}{|c||c|c|c|}
\hline
         & $\bm{s}^+$  & $\bm{b}^+$   & $\bm{b}^-$                                                             \\ \hline
Type I   & \begin{tabular}[c]{@{}c@{}}\phantom{x} \textbf{low} freq.\phantom{x}\\ (scaling)\end{tabular}  & \begin{tabular}[c]{@{}c@{}}\phantom{x} \textbf{mid} freq.\phantom{x}\\ (wavelet)\end{tabular} & \begin{tabular}[c]{@{}c@{}}\phantom{x} \textbf{high} freq.\phantom{x}\\ (wavelet)\end{tabular} \\ \hline
Type II  & \begin{tabular}[c]{@{}l@{}}\textbf{mid} freq.\\ (wavelet)\end{tabular}  & \begin{tabular}[c]{@{}l@{}}\textbf{low} freq.\\ (scaling)\end{tabular} & \begin{tabular}[c]{@{}l@{}}\textbf{high} freq.\\ (wavelet)\end{tabular} \\ \hline
Type III & \begin{tabular}[c]{@{}l@{}}\textbf{high} freq.\\ (wavelet)\end{tabular} & \begin{tabular}[c]{@{}l@{}}\textbf{low} freq.\\ (scaling)\end{tabular} & \begin{tabular}[c]{@{}l@{}}\textbf{mid} freq.\\ (wavelet)\end{tabular}  \\ \hline
\end{tabular}
\caption{Characterization of the three distinct possibilities of assigning a low/mid/high frequency range to the three coefficient sequences $\{ \bm{s}^ +,\bm{b}^ +,\bm{b}^- \}$ of a ternary unitary circuit. Note that Type I wavelets correspond to the site-centered circuit as depicted in Fig. \ref{fig:MultiTern}, while Types II-III correspond to the edge-centered circuit. Example wavelets of these types are presented in Fig. \ref{fig:AllTern}(a-c).} \label{tab:terntypes}
\end{table}

\subsubsection{Example 2: dilation $m=3$ wavelets with low/mid/high frequency components}
\begin{figure*} [!p]
  \begin{centering}
\includegraphics[width=12cm]{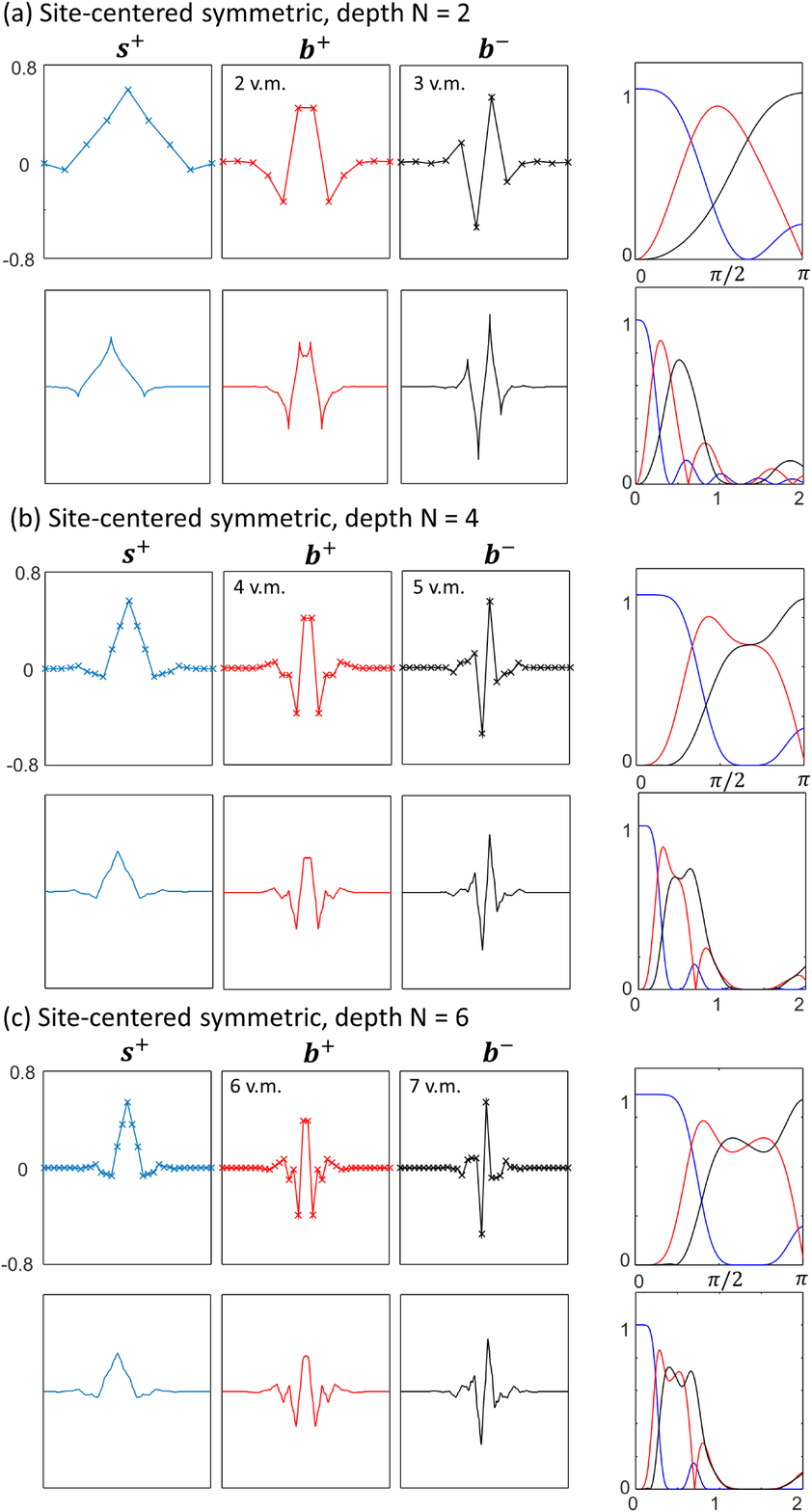}  \end{centering}
  \caption{Plots of the orthogonal wavelets given from site-centered symmetric ternary circuits of depths (a) $N = 2$, (b) $N = 4$ and (c) $N = 6$ with the angles $\theta_k$ specified in Tab. \ref{table:MaxTernary}. In each segment the scaling $\bm{s}^+$ and wavelet $\bm{b}^+$, $\bm{b}^-$ sequences are plotted, which possess the number of vanishing moments (v.m.) denoted, together with the scaling and wavelet functions in the continuum limit (windowed to include only the non-vanishingly small part of the functions). The fourth panels on the right depict the fourier transforms\cite{WaveBook3} of (upper) the scaling and wavelet sequences, and of (lower) the scaling and wavelet functions in the continuum limit.} \label{fig:MaxTern}
\end{figure*}

Choosing the angles $\{\theta_1,\theta_2,\ldots,\theta_N \}$ that parameterize the ternary circuit $\V$ to maximize the vanishing moments of the wavelets, as per Sect. \ref{sect:ex1}, has several notable deficiencies. One of these deficiencies if that the symmetric and anti-symmetric wavelets seen in Fig. \ref{fig:MaxTern} are not well separated when resolved in frequency space; ideally we want the three-channel filter bank represented by circuit $\V$ to possess well resolved low, mid, and high frequency components. A second deficiency is that the scaling functions in Fig. \ref{fig:MaxTern} still contain a significant high frequency component, which results in the corresponding wavelets not having a high degree of smoothness. In this Section we explore an alternative optimization criteria that resolves these deficiencies.

The alternative criteria we propose is that the circuit should be optimized such that some sequences $\bm{h} = [h_1,h_2,\ldots]$ have vanishing moments at high frequencies, 
\begin{equation}
\sum\limits_{r} {\Big( {(-1)^r (r^\alpha) {h_r}} \Big)}  = 0. \label{eq:s5e13}
\end{equation}
A three-channel wavelet filter bank with well resolved low/mid/high frequencies can be achieved by requiring that the scaling function and \emph{one} of the wavelets satisfy the (high frequency) orthogonality constraint of Eq. \ref{eq:s5e13} for some $\alpha=[0,1,\ldots ]$ (while also requiring that both wavelets satisfy the standard vanishing moment criteria of Eq. \ref{eq:s2e5}). Recall that the ternary unitary circuit has three associated sequences $\{\bm{s}^+, \bm{b}^+, \bm{b}^- \}$, as such there freedom as to how the sequences are associated to the low/mid/high frequency components (which also informs whether the bond-centered or edge-centered form of the ternary circuit is used, see Fig. \ref{fig:MultiTern}). Specifically, there are three choices (which we call Types I-III) that are compatible with the symmetries of the sequences, see Tab. \ref{tab:terntypes}.

In Tab. \ref{tab:ternangles} we present angles $\{\theta_1,\theta_2,\ldots,\theta_6 \}$ for ternary circuits of depth $N=6$ for each of the wavelet types I-III, which were again optimized numerically using Nelder-Mead minimization. The wavelet sequences were optimized to satisfy the (low frequency) orthogonality constraint of Eq. \ref{eq:s2e5} for $\alpha = [0,1,2]$, while the scaling and mid-frequency wavelet sequences were simultaneously optimized to satisfy the (high frequency) orthogonality constraint of Eq. \ref{eq:s5e13} for $\alpha = [0,1,2]$. Interesting to note is that the Type II wavelet has $\theta_2 =\theta_4 = \pi$ and $\theta_6 = 0$, which can be shown to imply that the sequence $\bm{b}^+ = [b^+_1, b^+_2, b^+_3, \ldots]$ is strictly zero for all even indexed entries, i.e. $b^+_r = 0$ for even $r$, which is compatible with its status as the mid-frequency wavelet. Furthermore, the final layer of $\theta_6 = 0$ gates could also be omitted, such that it could be regarded as a depth $N = 5$ ternary circuit that has an extra layer $U_\textrm{sw}$ of swap gates on the bottom. Notice that the deficiencies observed in the previous class of scale factor 3 wavelets, Fig. \ref{fig:MaxTern}, have been resolved; the new wavelets have high smoothness and are well separated into low/mid/high frequency components.  

%%%%%%%%%%%%%%%%%%%%%%%%%%%%%%%%%%%%%%%%%%%%%%%%%%%%%%
\begin{table}[!htb]
\centering
\begin{tabular}{|c||c|c|c|c|c|c|}
\hline 
       & \phantom{xxx} Type I \phantom{xxx}  & \phantom{xxx} Type II \phantom{xxx} & \phantom{xxx} Type III \phantom{xxx}  \\ \hline
$\theta_1$ &  0.072130476  & -0.261582176 &  0.072130476  \\
$\theta_2$ &  0.847695078  & $\pi$        & -0.847695078  \\
$\theta_3$ & -0.576099009  & 0.107465734  & -0.576099009  \\
$\theta_4$ & -0.591746629  & $\pi$        &  0.591746629  \\
$\theta_5$ &  0.673886987  & -0.461363266 &  0.673886987  \\
$\theta_6$ &  0.529449713  & 0            & -0.529449713  \\
 \hline
\end{tabular}
\caption{Angles $\theta_k$ parameterizing depth $N=6$ ternary unitary circuits which generate the wavelets depicted in Fig. \ref{fig:AllTern}.}
\label{tab:ternangles}
\end{table}
%%%%%%%%%%%%%%%%%%%%%%%%%%%%%%%%%%%%%%%%%%%%%%%%%%%%%%

%%%%%%%%%%%%%%%%%%%%%%%%%%%%%%%%%%%%%%%%%%%%%%%%%%%%%%
%%%%%%%%%%%%%%%%%%%%%%%%%%%%%%%%%%%%%%%%%%%%%%%%%%%%%%

%%%%%%%%%%%%%%%%%%%%%%%%%%%%%%%%%%%%%%%%%%%%%%%%%%%%%%
\begin{figure*} [!p]
  \begin{centering}
\includegraphics[width=12cm]{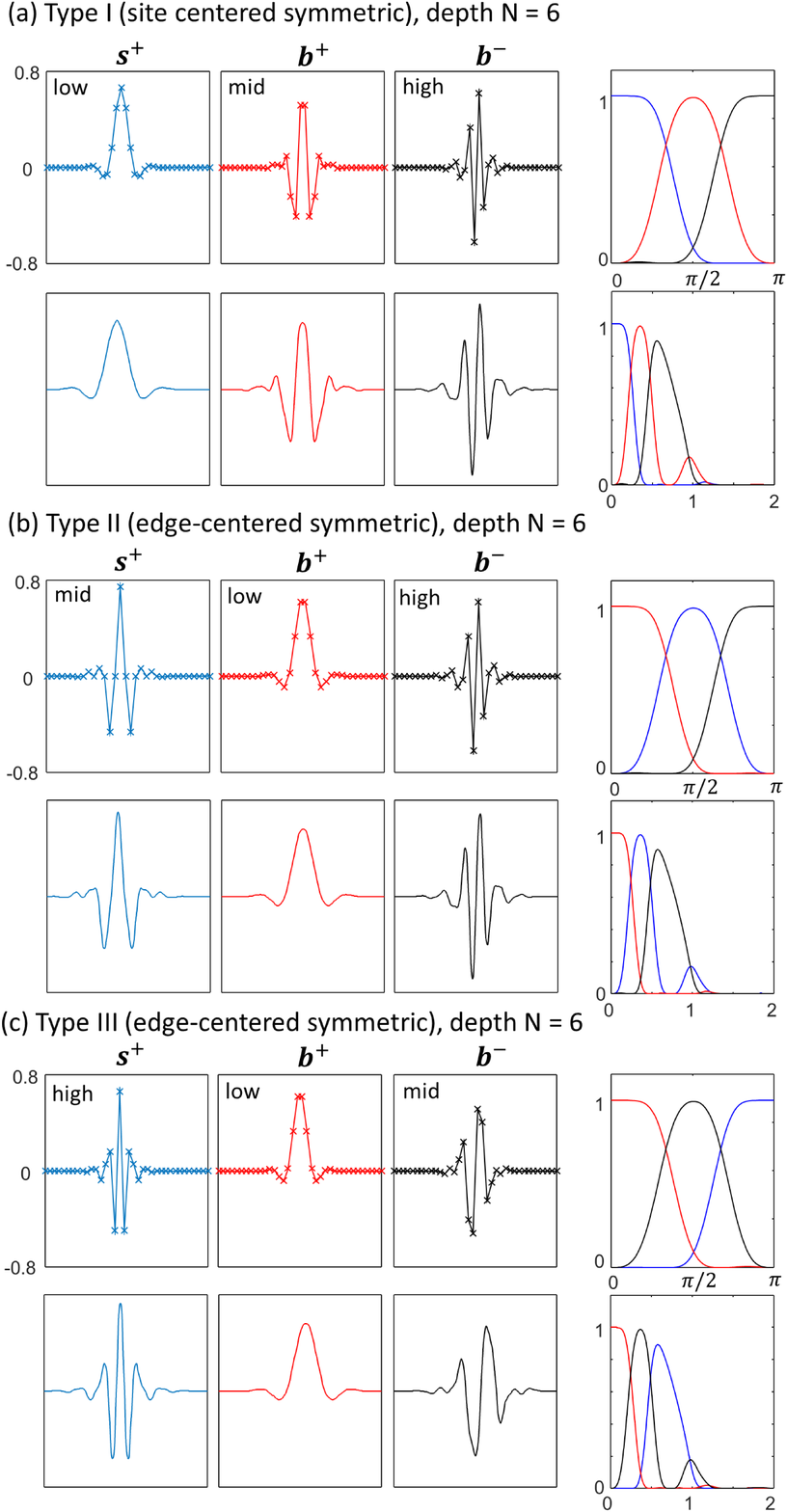}  \end{centering}
  \caption{Plots of the (a) Type I, (b) Type II and (c) Type III orthogonal wavelets, see Tab. \ref{tab:terntypes}, from depth $N = 6$ ternary circuits with angles as given in Tab. \ref{tab:ternangles}. In each segment the three sequences $\bm{s}^+$, $\bm{b}^+$, $\bm{b}^-$  are plotted, together with the scaling and wavelet functions in the continuum limit (windowed to include only the non-vanishingly small part of the functions). The fourth panels on the right depict the fourier transforms of (upper) the scaling and wavelet sequences, and of (lower) the scaling and wavelet functions in the continuum limit. In all cases, each of level of the wavelet transform is well resolved into low (scaling), mid (wavelet) and high (wavelet) frequency components.} \label{fig:AllTern}
\end{figure*}
%%%%%%%%%%%%%%%%%%%%%%%%%%%%%%%%%%%%%%%%%%%%%%%%%%%%%% 

%%%%%%%%%%%%%%%%%%%%%%%%%%%%%%%%%%%%%%%%%%%%%%%%%%%%%%
\begin{figure} [!htb]
  \begin{centering}
\includegraphics[width=8cm]{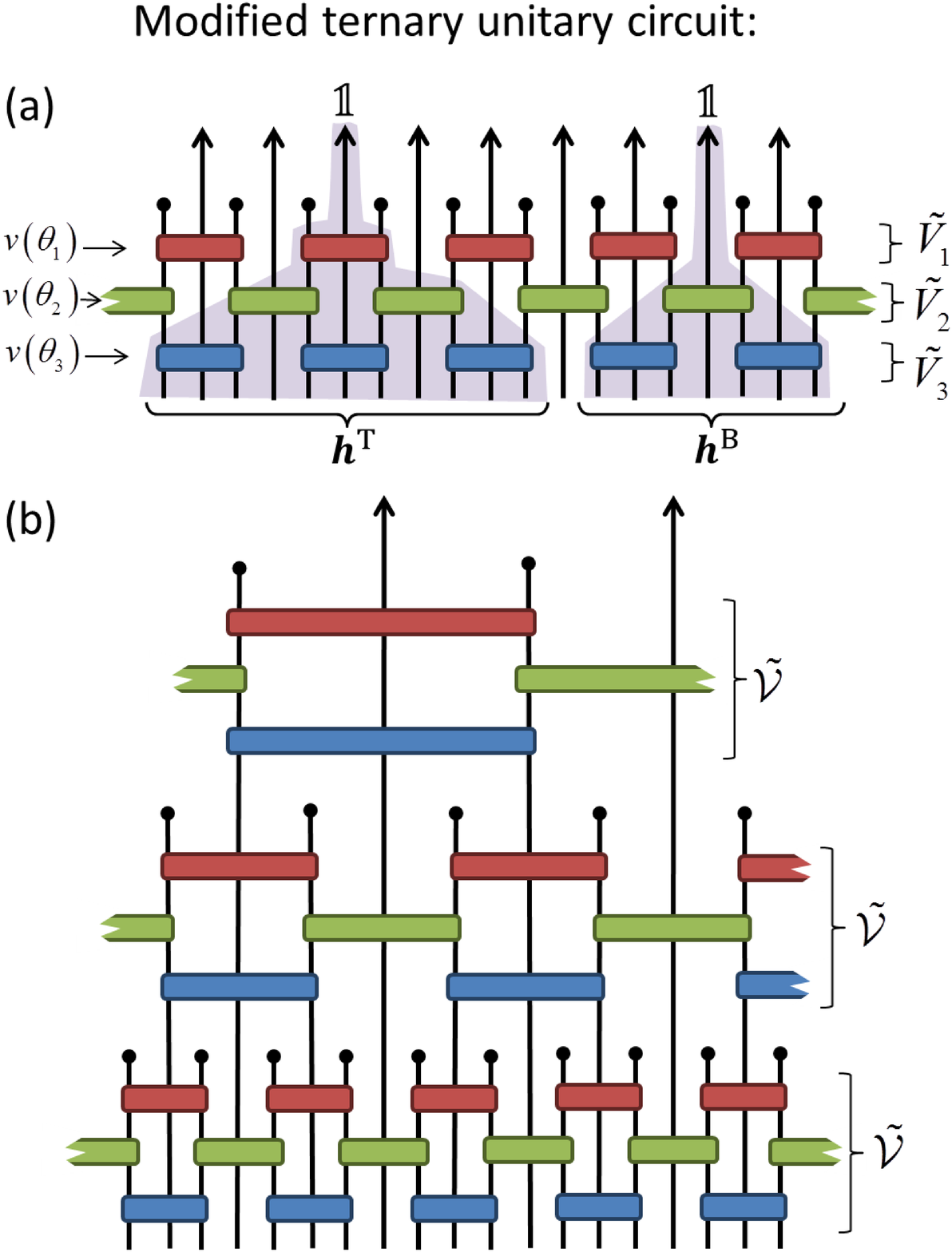}  \end{centering}
  \caption{(a) Depiction of a depth $N=3$ modified ternary circuit, which is built from reflection symmetric $3\times 3$ unitary matrices $v$ and parameterized by angles $\{ \theta_1, \theta_2, \theta_3\}$. Symmetric scaling sequences $\bm{h}^\textrm{T}$ and $\bm{h}^\textrm{B}$ are given from transforming unit vectors $\mathbbm{1}$ located on the central index of $v(\theta_1)$ and $v(\theta_2)$ respectively. (b) Multi-scale circuit formed from composition of depth $N=3$ modified ternary circuits.} \label{fig:MultiBinCircuit}
\end{figure}
%%%%%%%%%%%%%%%%%%%%%%%%%%%%%%%%%%%%%%%%%%%%%%%%%%%%%%

\subsection{Dilation $m=2$ symmetric multi-wavelet transform} \label{sect:multi}
An alternative solution to form symmetric orthogonal wavelets, other than from increasing the dilation factor $m>2$, is to use \emph{multiwavelets}, which possess more than one distinct scaling function (see for instance Refs. \onlinecite{Multi1,Multi2,Multi3,Multi4}). In this section we use the unitary circuit formalism to construct a family of dilation factor 2 multi-wavelets (noting that analogous constructions have previously been considered in the context of the MERA\cite{MERAbook}) where the two distinct scaling functions are each exactly reflection symmetric, and the two distinct wavelet functions are reflections of one another.

\subsubsection{Modified ternary unitary circuits}
Let $\tilde{\V}$ be a $4M \times 4M$ matrix; we say that $\tilde{\V}$ is a depth $N$ \emph{modified ternary circuit} if it can be decomposed as a product of $N$ layers,
%depicted in the circuit diagram of , or equivalently, can be written as,
\begin{equation}
\tilde{\V} = \tilde{V}_N \tilde{V}_{N-1} \ldots \tilde{V}_2 \tilde{V}_1, \label{eq:s6e1}
\end{equation}
see also Fig. \ref{fig:MultiBinCircuit}(a), where each layer $\tilde{V}_k$ is composed of a direct sum of $3\times 3$ reflection symmetric unitary matrices $v(\theta_k)$, as defined in Eq. \ref{eq:s5e4}, that are spaced by single site identities $\mathbb{I}$,
\begin{equation}
{\tilde{V}_k} = \ldots \oplus v(\theta_k) \oplus \mathbb{I} \oplus v(\theta_k) \oplus \mathbb{I} \oplus v(\theta_k) \oplus \mathbb{I} \oplus \ldots. \label{eq:s6e2}
\end{equation}
Additionally we require that each layer $\tilde{V}_k$ is offset by $r=2$ sites from the proceeding layer. Notice that the depth $N$ modified ternary circuit is parameterized by the set of angles $\{\theta_1, \theta_2,\ldots, \theta_N \}$.

The modified ternary circuit $\tilde{\V}$, which invariant under translations by four sites, may be interpreted as a four-channel filter bank. We label the four corresponding coefficient sequences $\bm h^T$, $\bm h^B$, $\bm g^L$ and $\bm g^R$, each of which is given by transforming a unit vector, i.e. as $\langle \tilde{\V} \times \mathbbm 1_r \rangle$. The sequences possesses the following properties,
\begin{itemize}
\item $\bm{h}^ T$ is a site-centered symmetric sequence of (odd) length $4N-1$ elements (i.e. that is symmetric w.r.t. reflections about the central element), and is given from transforming the unit vector $\mathbbm 1$ located at the center of $v(\theta_1)$, see Fig. \ref{fig:MultiBinCircuit}(a). 
\item $\bm{h}^ B$ is a site-centered symmetric sequence of (odd) length $4N-5$ elements (i.e. that is symmetric w.r.t. reflections about the central element), and is given from transforming the unit vector $\mathbbm 1$ located at the center of $v(\theta_2)$, see Fig. \ref{fig:MultiBinCircuit}(a). 
\item $\bm{g}^ L$ and $\bm{g}^ R$ are length $4N-1$ sequences given from transforming the unit vector $\mathbbm 1$ located at the left and right index of $v(\theta_2)$ respectively, and are related to each other by spatial reflection.
\end{itemize}
A multi-scale circuit can be formed by composition of modified ternary circuits, $\tilde{\V} \circ \tilde{\V} \circ \tilde{\V} \circ \ldots$, where both $\bm h^T$ and $\bm h^B$ are treated as scaling sequences, see also Fig. \ref{fig:MultiBinCircuit}(b), which in turn represents the multiwavelet transform of dilation factor $m=2$.

%%%%%%%%%%%%%%%%%%%%%%%%%%%%%%%%%%%%%%%%%%%%%%%%%%%%%%
\begin{table}[htb]
\centering
\begin{tabular}{|c||c|c|}
\hline
       & $N = 4$    & $N = 9$              \\ \hline
$\theta_1$ & \phantom{x} 0.161653803 \phantom{x} & \phantom{x} 0.190056742 \phantom{x}  \\ 
$\theta_2$       & 0.389671265 & 0.438258716  \\ 
$\theta_3$       & 0.395157917 & 0.389371447 \\ 
$\theta_4$       & 0.167734951 & -0.938456988\\ 
$\theta_5$       &   -         & 0.319668509  \\ 
$\theta_6$       &    -        & 0.875926194  \\ 
$\theta_7$       &    -        & -1.048336048 \\ 
$\theta_8$       &    -        & 0.314028313  \\ 
$\theta_9$       &    -        & 0.843502411  \\ \hline
\end{tabular}
\caption{Angles $\theta_k$ parameterizing depth $N = \{4,9\}$ modified ternary circuits, see Fig. \ref{fig:MultiBinCircuit}, that yield the wavelets depicted in Fig. \ref{fig:MultiBinWave}.} \label{tab:MultiBin}
\end{table}
%%%%%%%%%%%%%%%%%%%%%%%%%%%%%%%%%%%%%%%%%%%%%%%%%%%%%%

%%%%%%%%%%%%%%%%%%%%%%%%%%%%%%%%%%%%%%%%%%%%%%%%%%%%%%
\begin{figure} [!htb]
  \begin{centering}
\includegraphics[width=8cm]{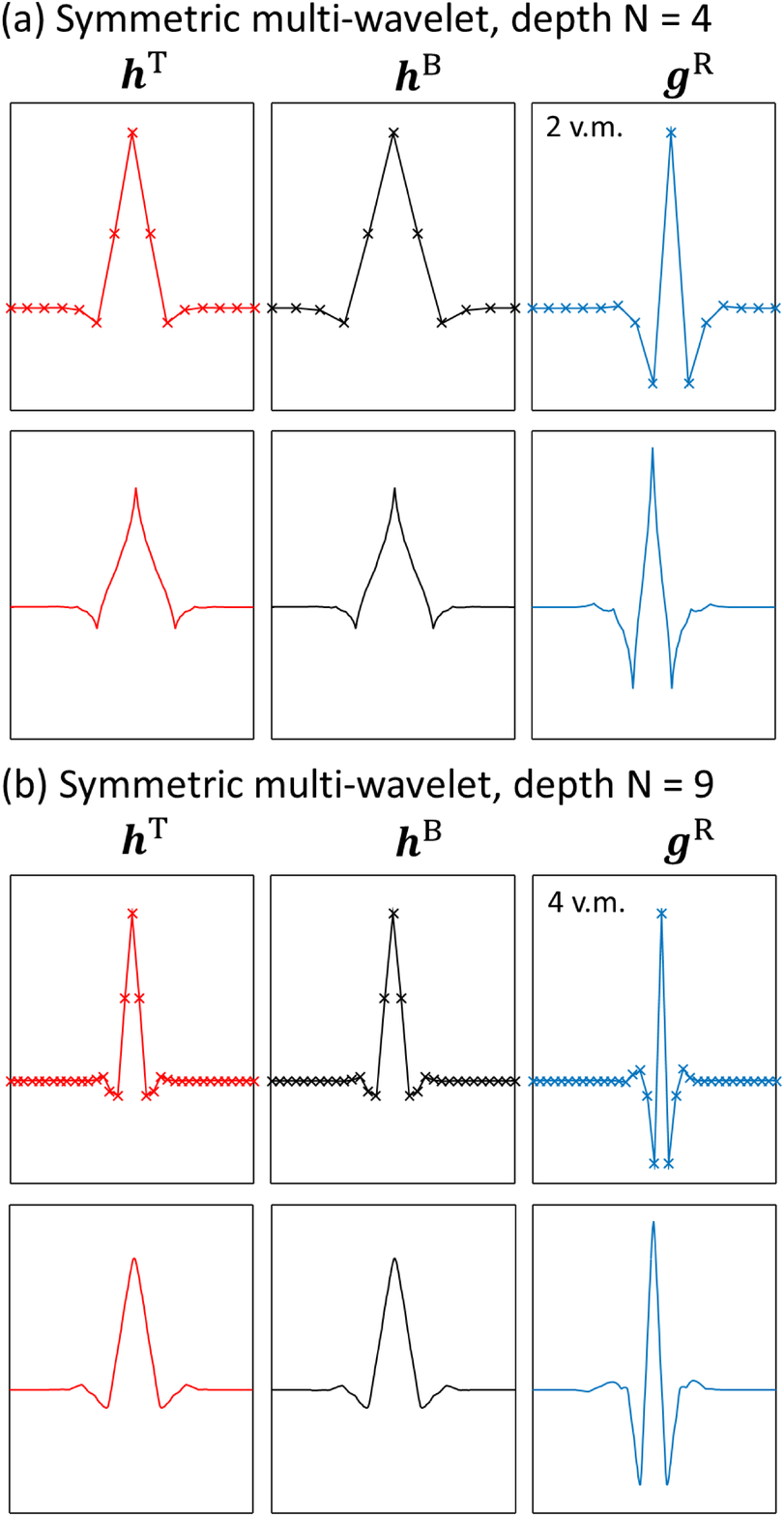}  \end{centering}
  \caption{Plots of orthogonal multi-wavelets from (a) depth $N=4$ and (b) depth $N=9$ modified ternary circuits with angles $\theta_k$ as given in Tab. \ref{tab:MultiBin}. The top three panels of each group denote the (exactly symmetric) scaling sequences $\bm{h}^\textrm{T}$ and $\bm{h}^\textrm{B}$, and the wavelet sequence $\bm{g}^\textrm{R}$ (where $\bm{g}^\textrm{L}$, not shown, is the mirror of $\bm{g}^\textrm{R}$). The bottom three panels of each group depict the scaling functions and wavelets in the continuum limit (windowed to include only the non-vanishingly small part of the functions).} \label{fig:MultiBinWave}
\end{figure}
%%%%%%%%%%%%%%%%%%%%%%%%%%%%%%%%%%%%%%%%%%%%%%%%%%%%%%  

\subsubsection{Example 3: dilation $m=2$ symmetric multi-wavelets} 
There are many possible criteria for optimization of the angles $\{\theta_1, \theta_2,\ldots, \theta_N \}$ associated to the depth $N$ modified ternary circuit $\tilde{\V}$. One choice could be to optimize to maximize the number of vanishing moments that the wavelets possess. For instance, if angles $\theta_k$ were chosen such that the wavelet sequences $\bm g^L$ and $\bm g^R$ possessed vanishing moments, see Eq. \ref{eq:s2e5}, for $\alpha = [0,1,\ldots,P - 1]$ and the scaling sequences $\bm h^T$ and $\bm h^B$ had vanishing moments for $\alpha = [1,\ldots,P - 1]$, then the family of multi-wavelets given from composition of $\tilde{\V}$ would possess $P$ vanishing moments. Using this criteria we have found solutions for circuits depth $N = 3$ with $P = 2$ vanishing moments, and for depth $N = 6$ with $P = 4$ vanishing moments, although we do not present these results here.

Instead, in Tab. \ref{tab:MultiBin} we present solutions for depth $N = 4$ with $P = 2$ vanishing moments, and for depth $N = 9$ with $P = 4$ vanishing moments. Here the extra degrees of freedom (resulting from use of larger depth $N$ circuits than is necessary to achieve the desired number of vanishing moments) have been used to minimize the difference between the two distinct scaling sequences $\bm h^T$ and $\bm h^B$ of the multiwavelet transform, which also imposes that the wavelet sequences $\bm{g}^ L$ and $\bm{g}^ R$ are each individually close to being reflection symmetric. This was achieved using the Nelder-Mead algorithm to numerically optimize the angles $\{\theta_1,\theta_2,\ldots,\theta_N \}$ as to minimize the magnitude of the desired scaling and wavelet sequence moments, as usual, but where a small term was also included in the figure of merits that penalized the difference between the two distinct scaling sequences. The term used was of the form $c \Omega$, where
\begin{equation}
\Omega = || \bm{h}^T - \bm{h}^B ||, \label{eq:s6e2b}
\end{equation}
and scalar $c$ was a small fixed parameter. Note that the sequence $\bm{h}^B$ in Eq. \ref{eq:s6e2b} has been padded by two zero elements at the start and end of the sequence, such that the padded sequence is of the same length as $\bm{h}^T$.  

The resulting wavelets are plotted in Fig. \ref{fig:MultiBinWave}, which are seen to closely resemble the corresponding coiflets with $P = 2$ and $P = 4$ vanishing moments. In both examples only very small differences between the two scaling sequences are realized; for the depth $N=4$ example we have $|| \bm{h}^T - \bm{h}^B || = 0.0005$, while for the depth $N=9$ example we have $|| \bm{h}^T - \bm{h}^B || = 0.0048$. In comparison with coiflets, where the scaling functions are only approximately reflection symmetric, here we have achieved exact symmetry at the expense of now having two distinct (yet almost identical) scaling functions. 

\subsection{Dilation $m=4$ symmetric wavelets} \label{sect:dilation4}
In this section we construct an example family of exactly reflection symmetric and antisymmetric wavelets of dilation factor $m=4$ using a unitary circuit formalism. The purpose of this example is to demonstrate an alternate construction of symmetric wavelets that is not based upon the individually symmetric $3\times 3$ unitary matrices $v(\theta)$ as considered in Sects. \ref{sect:dilation3} and \ref{sect:multi}. Instead, the circuit that represents the symmetric wavelet transform is built from $2\times 2$ unitary matrices. 

%%%%%%%%%%%%%%%%%%%%%%%%%%%%%%%%%%%%%%%%%%%%%%%%%%%%%%
\begin{figure} [!htb]
  \begin{centering}
\includegraphics[width=8.5cm]{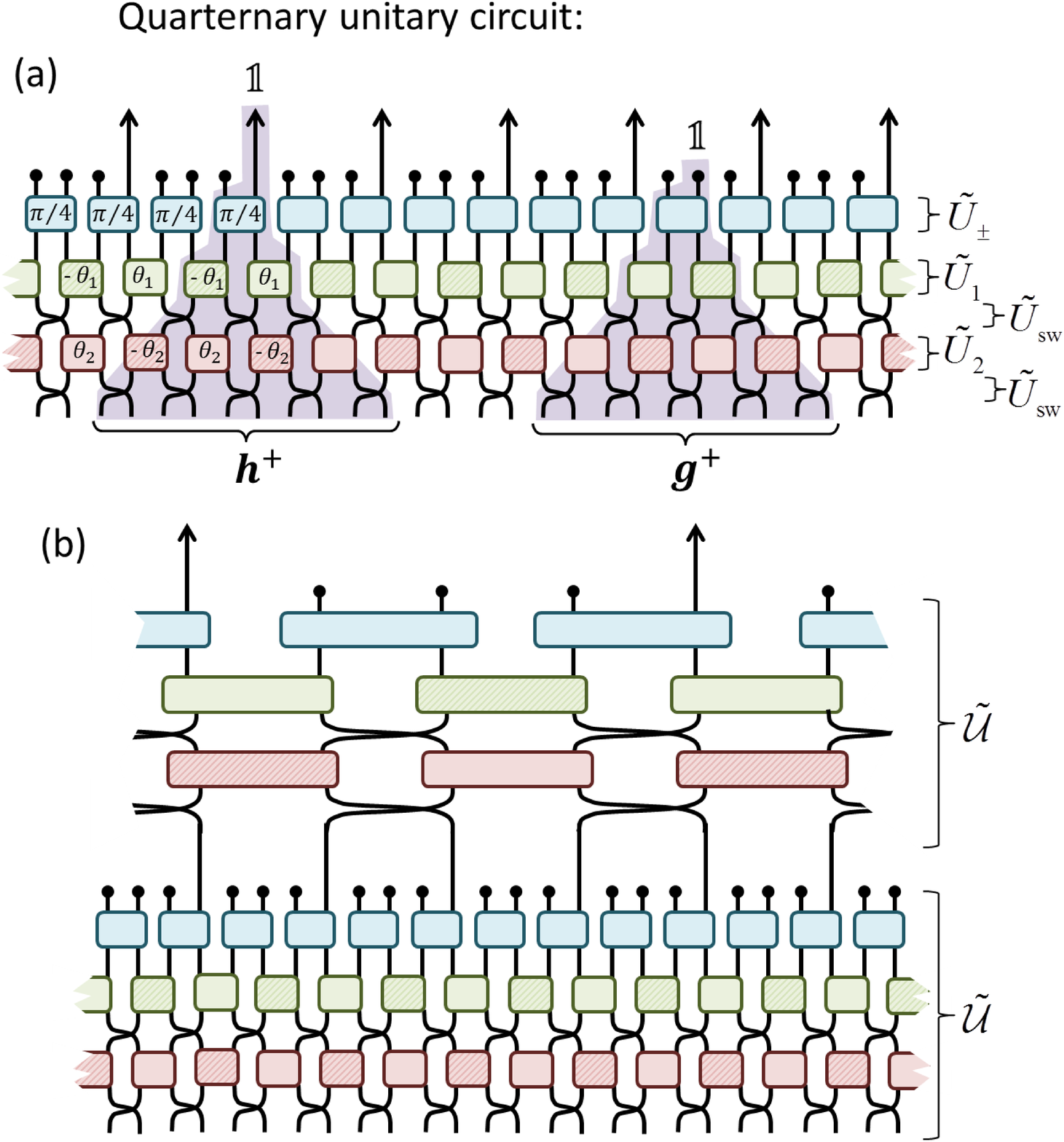}  \end{centering}
\caption{(a) Depiction of a depth $N=2$ quarternary circuit, which is built from $2\times 2$ unitary matrices $u$ and parameterized by angles $\{ \theta_1, \theta_2\}$ as labelled. Symmetric sequences $\bm{g}^+$ and $\bm{h}^+$ are given from transforming unit vectors $\mathbbm{1}$ located on a right-side index of an top level gate as indicated (while antisymmetric sequences $\bm{g}^-$ and $\bm{h}^-$ are given from transforming unit vectors $\mathbbm{1}$ located on a left-side index). (b) Multi-scale circuit formed from composition of depth $N=3$ quarternary circuits.} \label{fig:QuadCircuit}
\end{figure}
%%%%%%%%%%%%%%%%%%%%%%%%%%%%%%%%%%%%%%%%%%%%%%%%%%%%%%

\subsubsection{Quaternary circuits}
Let $\tilde{\U}$ be a $4M \times 4M$ matrix for positive integer $M$; we say that $\tilde{\U}$ is a depth $N$ \emph{quaternary unitary circuit} if it can be decomposed into $N$ sublayers $\tilde{U}_k$, each of which is a direct sum of $2\times 2$ unitary matrices $u$,
\begin{equation}
\tilde{U}_k = \ldots \oplus u(\theta_k) \oplus u(-\theta_k) \oplus u(\theta_k) \oplus u(-\theta_k) \oplus \ldots, \label{eq:s7e1}
\end{equation}
where unitaries $u$ are parameterized by a rotation angle $\theta_k$, see Eq. \ref{eq:s1e3}, which alternates in sign with position, see also Fig. \ref{fig:QuadCircuit}(a). Layers $\tilde{U}_k$ of the quaternary unitary circuit $\tilde{\U}$ are inter-spaced with layers $\tilde U_\textrm{sw}$ of swap gates,
\begin{equation}
{\tilde U}_\textrm{sw} = \ldots \oplus u_\textrm{sw} \oplus u_\textrm{sw} \oplus u_\textrm{sw} \oplus u_\textrm{sw} \ldots, \label{eq:s7e2}
\end{equation}
with $u_\textrm{sw}$ the two site swap gate as defined in Eq. \ref{eq:s5e2}, and the quaternary circuit also includes a top sublayer $\tilde U_{\pm}$ composed of $2\times 2$ unitary gates $u(\pi/4)$, 
\begin{equation}
{\tilde U}_{\pm} = \ldots \oplus \; u(\pi/4) \; \oplus \; u(\pi/4) \; \oplus \; u(\pi/4) \ldots, \label{eq:s7e3}
\end{equation}
Notice that the depth $N$ quaternary unitary circuit is parameterized by the set of $N$ angles $\{\theta_1, \theta_2, \ldots, \theta_N \}$.

The quaternary unitary circuit, which is invariant w.r.t translations of four sites, may be interpreted as a four-channel filter bank. The four corresponding coefficient sequences $\{ \bm{h}^+, \bm{h}^-, \bm{g}^+, \bm{g}^- \}$, each given from transforming the unit vector ${\mathbbm 1}_r$ on a non-equivalent site, have the following properties:
\begin{itemize}
\item $\bm{h}^ +$ is a edge-centered symmetric sequence of (even) length $4N+2$ elements, given from transforming ${\mathbbm 1}_r$ with $r=0,4,8,\ldots$.%: $\mod(r,4) = 0$.
\item $\bm{h}^ -$ is a edge-centered antisymmetric sequence of (even) length $4N+2$ elements, given from transforming ${\mathbbm 1}_r$ with $r=1,5,9,\ldots$.%: $\mod(r,4) = 1$.
\item $\bm{g}^ +$ is a edge-centered symmetric sequence of (even) length $4N+2$ elements, given from transforming ${\mathbbm 1}_r$ with $r=2,6,10,\ldots$.%: $\mod(r,4) = 2$.
\item $\bm{g}^ -$ is a edge-centered antisymmetric sequence of (even) length $4N+2$ elements, given from transforming ${\mathbbm 1}_r$ with $r=3,7,11,\ldots$.%: $\mod(r,4) = 3$.
\end{itemize}
The symmetry of the coefficient sequences can be understood by recalling that $\mathcal R \left( u(\theta) \right) = u(-\theta)$, where $\mathcal R$ denotes spatial reflection, which implies that $\left( u(\theta)\oplus u(-\theta) \right)$ constitutes a $4\times 4$ unitary matrix that is reflection symmetric.

A multi-scale circuit, which represents a family of wavelet transform of dilation factor $m=4$, is formed by composition $\tilde{\U} \circ \tilde{\U} \circ \tilde{\U} \circ \ldots$ of quaternary circuits, where $\bm h^+$ is treated as the scaling sequence, see also Fig. \ref{fig:QuadCircuit}(b). 

\subsubsection{Example 4: dilation $m=4$ symmetric wavelets}
\begin{figure} [!htb]
  \begin{centering}
\includegraphics[width=8cm]{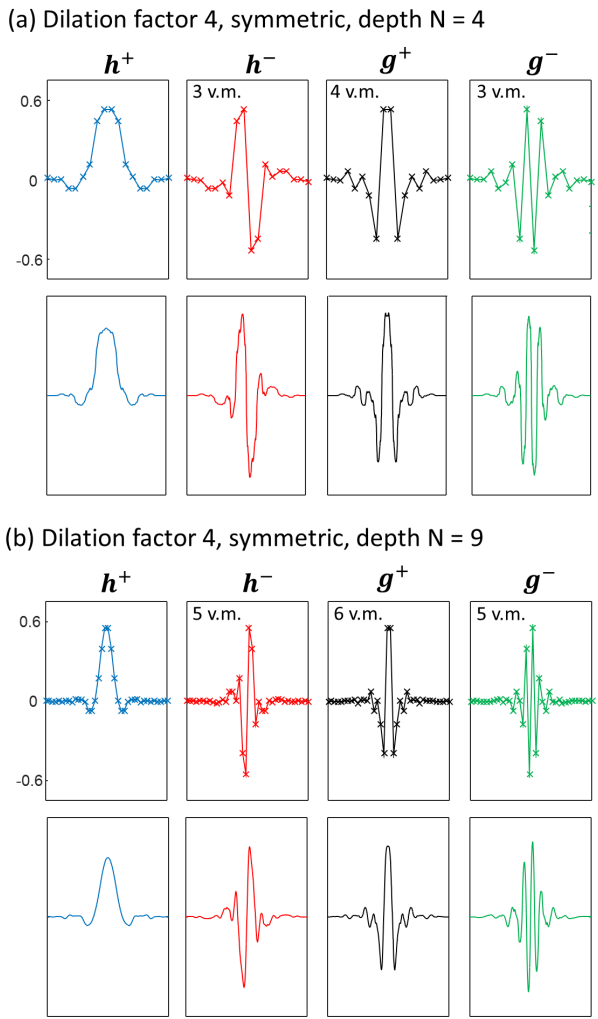}  \end{centering}
  \caption{Plots of dilation $m=4$ orthogonal wavelets from (a) depth $N=4$ and (b) depth $N=9$ quarternary circuits with angles $\theta_k$ as given in Tab. \ref{tab:Quad}. The top three panels of each group denote the (exactly symmetric) scaling sequence $\bm{h}^+$ and wavelet sequences $\bm{h}^-$, $\bm{g}^+$, $\bm{h}^-$, which possess the number of vanishing moments (v.m.) as indicated. The bottom three panels of each group depict the scaling functions and wavelets in the continuum limit (windowed to include only the non-vanishingly small part of the functions).} \label{fig:QuadWave}
\end{figure}

For a depth $N$ quaternary circuit, we optimize the free angles $\{\theta_1, \theta_2,\ldots, \theta_N \}$ numerically using the Nelder-Mead algorithm to give wavelet sequences $\{ \bm{h}^ -, \bm{g}^ +, \bm{g}^ - \} $ with as many vanishing moments as found to be possible. Two sets of example angles, one for depth $N = 4$ and the other for depth $N = 9$, are given in Tab. \ref{tab:Quad}. The depth $N = 4$ circuit yields wavelets with vanishing moments for orders $\alpha = [0,1,2]$, while the depth $N = 9$ circuit yields wavelets with vanishing moments for orders $\alpha = [0,1,2,3,4]$. These example wavelets, shown in Fig. \ref{fig:QuadWave}, display a relatively high smoothness while also maintaining a small effective support, which suggests they may perform well in perform well in practical applications. 

%%%%%%%%%%%%%%%%%%%%%%%%%%%%%%%%%%%%%%%%%%%%%%%%%%%%%%
\begin{table}[htb]
\centering
\begin{tabular}{|c||c|c|}
\hline
       & $N = 4$    & $N = 9$            \\ \hline
$\theta_1$       & \phantom{x} -1.229229752 \phantom{x} & \phantom{x} 2.807164665 \phantom{x} \\ 
$\theta_2$       &  0.102977579 & -0.473049662   \\ 
$\theta_3$       &  1.918752569 & -1.482553131  \\ 
$\theta_4$       & -0.198852925 & -1.922365069  \\
$\theta_5$       &    -       & -3.091128284  \\ 
$\theta_6$       &     -      &  2.257247480  \\ 
$\theta_7$       &      -     &  2.554082219  \\ 
$\theta_8$       &       -    & -0.492838196  \\ 
$\theta_9$       &        -   & -0.633172753  \\ \hline
\end{tabular}
\caption{Angles parameterizing depth $N={4,9}$ quarternary circuits, as depicted in Fig. \ref{fig:QuadCircuit}, which produce the wavelets depicted in Fig. \ref{fig:QuadWave}.} \label{tab:Quad}
\end{table}
%%%%%%%%%%%%%%%%%%%%%%%%%%%%%%%%%%%%%%%%%%%%%%%%%%%%%%

%%%%%%%%%%%%%%%%%%%%%%%%%%%%%%%%%%%%%%%%%%%%%%%%%%%%%%
%%%%%%%%%%%%%%%%%%%%%%%%%%%%%%%%%%%%%%%%%%%%%%%%%%%%%%

\subsection{Boundary wavelets} \label{sect:boundary}
In this section we explore the use of unitary circuits in the construction of boundary wavelets\cite{Bound1,Bound2,Bound3,Bound4}, using a similar approach to the boundary MERA developed in the context of studying quantum spin chains with open boundaries\cite{BoundMERA1,BoundMERA2}.

%%%%%%%%%%%%%%%%%%%%%%%%%%%%%%%%%%%%%%%%%%%%%%%%%%%%%%
\begin{figure} [!htb]
  \begin{centering}
\includegraphics[width=8cm]{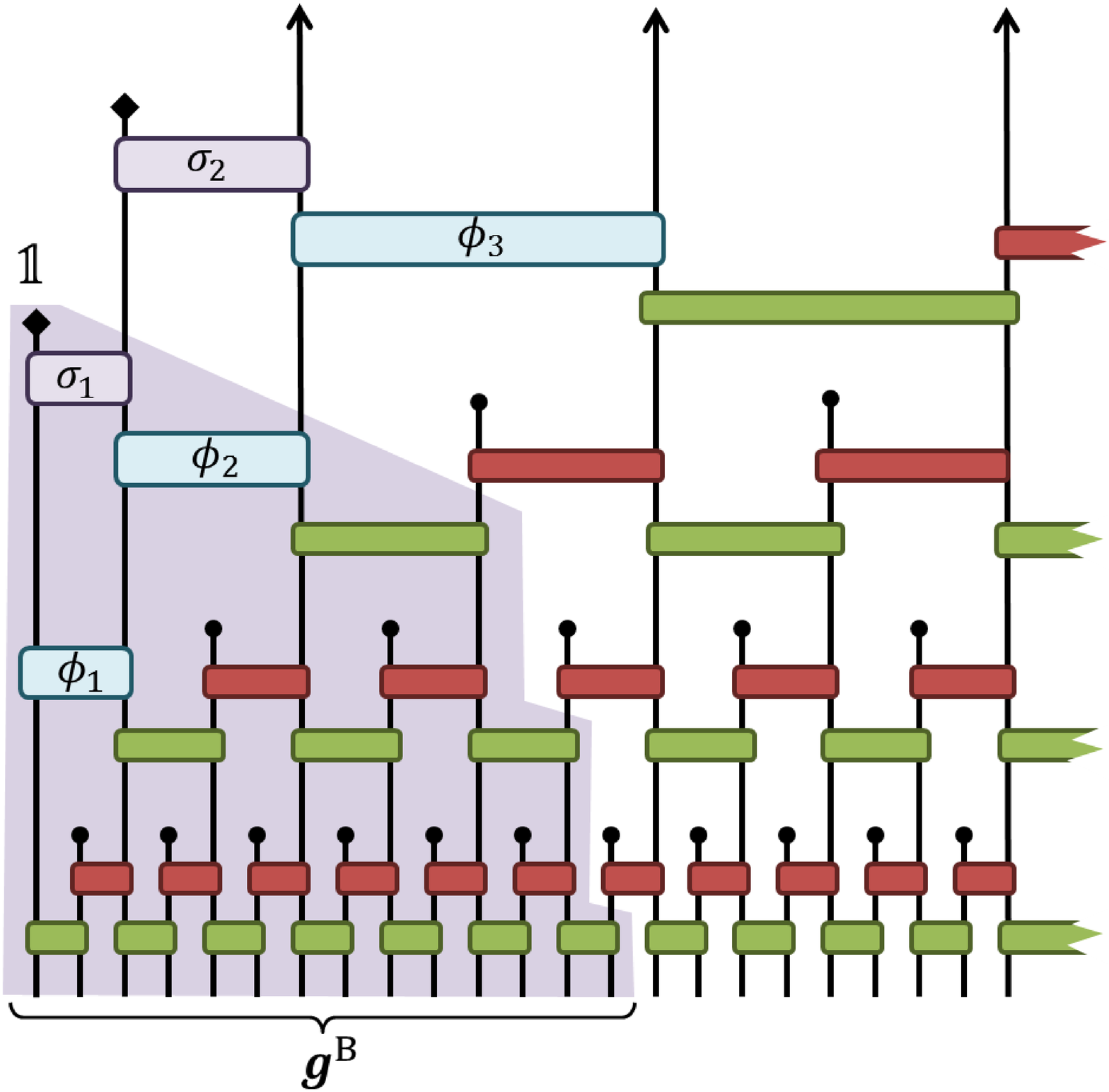}  \end{centering}
  \caption{Depiction of a multi-scale circuit (formed from composition of depth $N=2$ binary unitary circuits) with an open boundary on the left, where a double layer of (scale dependent) unitary gates $u(\phi_z)$ and $u(\sigma_z)$ is introduced. Boundary wavelets $\bm{g}^\textrm{B}$ are given by transforming the unit vector $\mathbbm{1}$ located on a boundary index as indicated. The angles $\phi_z$ and $\sigma_z$ are chosen to ensure the boundary wavelets each have two vanishing moments.} \label{fig:BoundCircuit}
\end{figure}
%%%%%%%%%%%%%%%%%%%%%%%%%%%%%%%%%%%%%%%%%%%%%%%%%%%%%%

%%%%%%%%%%%%%%%%%%%%%%%%%%%%%%%%%%%%%%%%%%%%%%%%%%%%%%%%%%%%%%%%%%%%%%%%%%%%%%%
\begin{table}[tbh]
%\begin{tabular}{|l||l|l|l|l|l|}
\begin{tabular}{|c||c|c|}
\hline
& \phantom{xxxxx}$\phi_z$\phantom{xxxxx}  &\phantom{xxxxx} $\sigma_z$\phantom{xxxxx}   \\ \hline
$z=1 $ &	 0.615479708 &  0.261157410 \\
$z=2 $ &     0.713724378 &  0.316335000 \\
$z=3 $ &     0.752040089 &  0.339836909 \\
$z=4 $ &     0.769266332 &  0.350823961 \\
$z=5 $ &     0.777461322 &  0.356148400 \\
$z=6 $ &     0.781461114 &  0.358770670 \\
$z=7 $ &     0.783437374 &  0.360072087 \\
$z=8 $ &     0.784419689 &  0.360720398 \\
$z=9 $ &     0.784909404 &  0.361043958 \\
$z=10 $ &    0.785153903 &  0.361205590 \\
$z=11 $ &    0.785276063 &  0.361286369 \\
$z=12 $ &    0.785337120 &  0.361326749 \\					
\hline
\end{tabular}
\caption{Angles $\phi_z$ and $\sigma_z$ parameterizing the boundary unitary gates at scale $z$, see Fig. \ref{fig:BoundCircuit}, of the boundary D4 Daubechies wavelets.}
\label{table:Bound}
\end{table}
%%%%%%%%%%%%%%%%%%%%%%%%%%%%%%%%%%%%%%%%%%%%%%%%%%%%%%%%%%%%%%%%%%%%%%%%%%%%%%%	

%%%%%%%%%%%%%%%%%%%%%%%%%%%%%%%%%%%%%%%%%%%%%%%%%%%%%%
\begin{figure} [!tbh]
  \begin{centering}
\includegraphics[width=8cm]{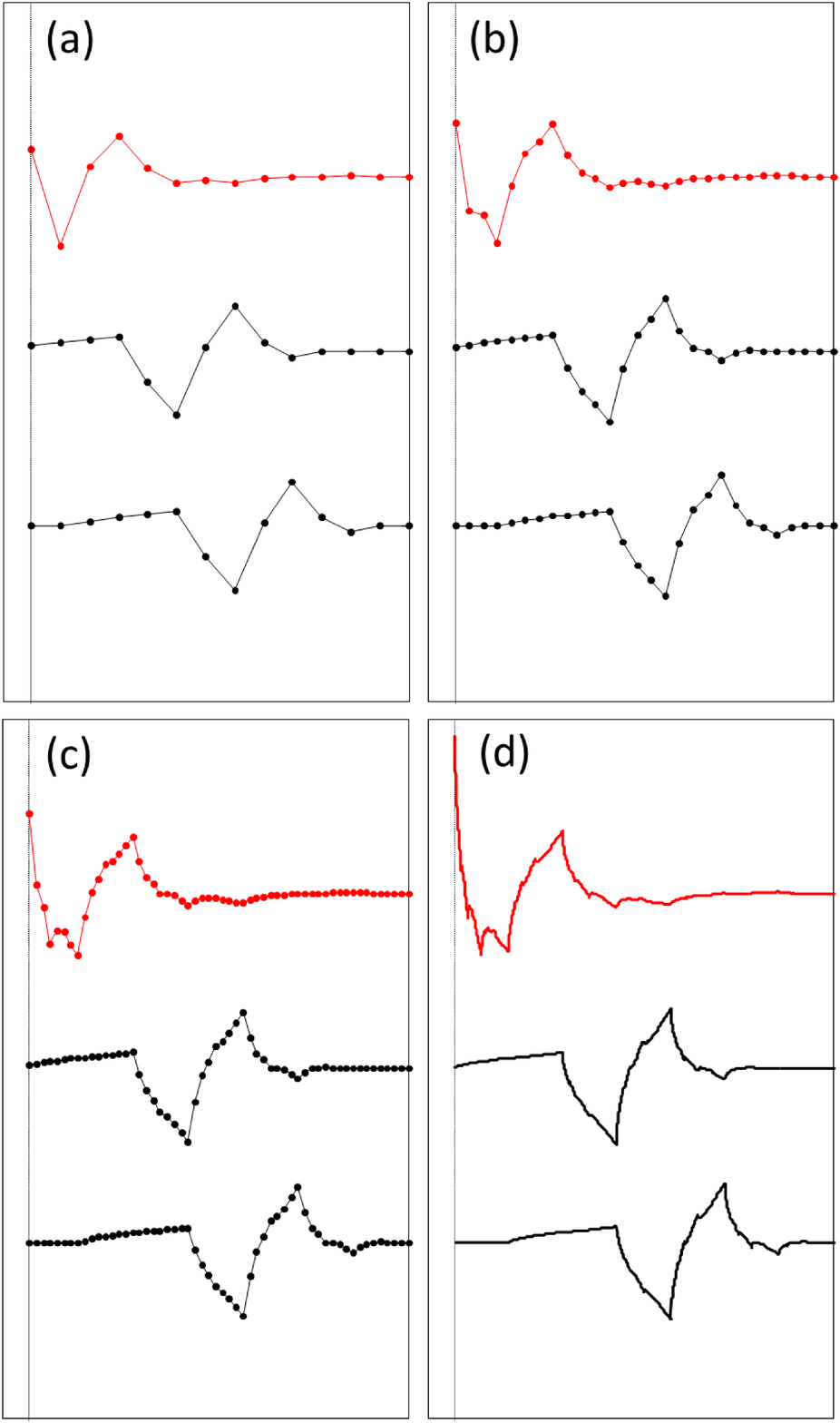}  \end{centering}
  \caption{Plots of the orthogonal boundary wavelet (red) and two of the bulk D4 Daubechies wavelets at the same scale (black) from the boundary unitary circuit depicted in Fig. \ref{fig:BoundCircuit}, with the angles as specified in Tab. \ref{table:Bound} for (a) scale $z=1$, (b) scale $z=2$, (c) scale $z=3$ and (d) the continuum limit. All wavelets are orthogonal and have two vanishing moments.} \label{fig:BoundWave}
\end{figure}
%%%%%%%%%%%%%%%%%%%%%%%%%%%%%%%%%%%%%%%%%%%%%%%%%%%%%%

\subsubsection{Example 5: boundary wavelets for D4 Daubechies wavelets}
Here we provide an example the use of a unitary circuit in the construction of boundary wavelets for the specific case of the D4 Daubechies wavelets, although the general strategy we employ for constructing the boundary wavelets can easily be extended not only to higher order Daubechies wavelets, but also to other wavelet families. The boundary wavelets we construct are orthogonal and, as with the D4 Daubechies wavelets, possess two vanishing moments.

The boundary wavelets are parameterized using a \emph{boundary unitary circuit}, see Fig. \ref{fig:BoundCircuit}, which is constructed as follows. Let $\mathcal L$ be a semi-infinite lattice of sites $r = [1,2,3,\ldots]$, i.e. with an open boundary on the left. We apply as much of the multi-scale circuit corresponding to the D4 wavelets, i.e. composed of depth $N=2$ binary unitary circuits with angles $\theta_1 = 5\pi/12$ and $\theta_2 = \pi/6$ (see Sect. \ref{sect:D4}), that can be supported on $\mathcal L$. Then a double layer of scale-dependent $2\times 2$ unitary gates, parameterized by angles $\{\phi_1, \phi_2, \phi_3, \ldots \}$ and  $\{\sigma_1, \sigma_2, \sigma_3, \ldots \}$, where subscripts here denote scale $z$, is introduced on the boundary of the multi-scale circuit.

Starting from the lowest scale the angles $\phi_1$ and $\sigma_1$ are fixed such that the corresponding boundary wavelet $\bm{g}^\textrm{B}$ at that scale, as depicted in Fig. \ref{fig:BoundCircuit}(a), has two vanishing moments. Here the angles $\phi_1$ and $\sigma_1$ can be found deterministically using a algorithm similar to the circuit construction algorithm discussed in Sect. \ref{sect:algorithm}. We then apply the algorithm iteratively, scale-by-scale, to fix remaining angles $\{\phi_2, \phi_3, \ldots \}$ and  $\{\sigma_2, \sigma_3, \ldots \}$ such that all boundary wavelets have two vanishing moments. The resulting angles are given in Tab. \ref{table:Bound}, and are seen to converge to the values $\phi_z = \pi/4$ and $\sigma_z = 0.361367123906708$ in the large scale limit, $z\rightarrow \infty$.

The resulting boundary wavelets are plotted in Fig. \ref{fig:BoundWave} over several different scales $z$. Notice that the double layer of unitary gates was used in the boundary circuit of Fig. \ref{fig:BoundWave} in order to provide two degrees of freedom $\phi_z$ and $\sigma_z$ for each boundary wavelet (which could then be chosen in order to endow each boundary wavelet with two vanishing moments). More generally additional layers of boundary unitary gates could be included in order to generate wavelets with higher vanishing moments. 

\subsection{Biorthogonal wavelets} \label{sect:biorthog}
Biorthogonal wavelets, such as the CDF wavelets introduced in Ref. \onlinecite{Daub3}, follow from removing the orthogonality constraint on wavelets while still requiring perfect reconstruction. For a given number of vanishing moments, biorthogonal wavelets can possess smaller support than orthogonal wavelets, and can be exactly symmetric even for dilation factor $m=2$. These properties make biorthogonal wavelets useful in practical applications such as image compression, where they are employed in the JPEG2000 format \cite{JPEG}.

In this section we briefly discuss the use of a circuit formalism in the construction of biorthogonal wavelets. First we introduce the notion of invertible circuits which are built from many copies of local invertible matrices, and follow from relaxing the unitary constraint on the circuit formalism. As with the case of the unitary circuits explored in this manuscript, many different forms of invertible circuit could be considered; here the example we provide leads to a novel family of (edge-centered) antisymmetric biorthogonal wavelets.

\subsubsection{Binary invertible circuits}
\begin{figure} [!tbh]
  \begin{centering}
\includegraphics[width=8cm]{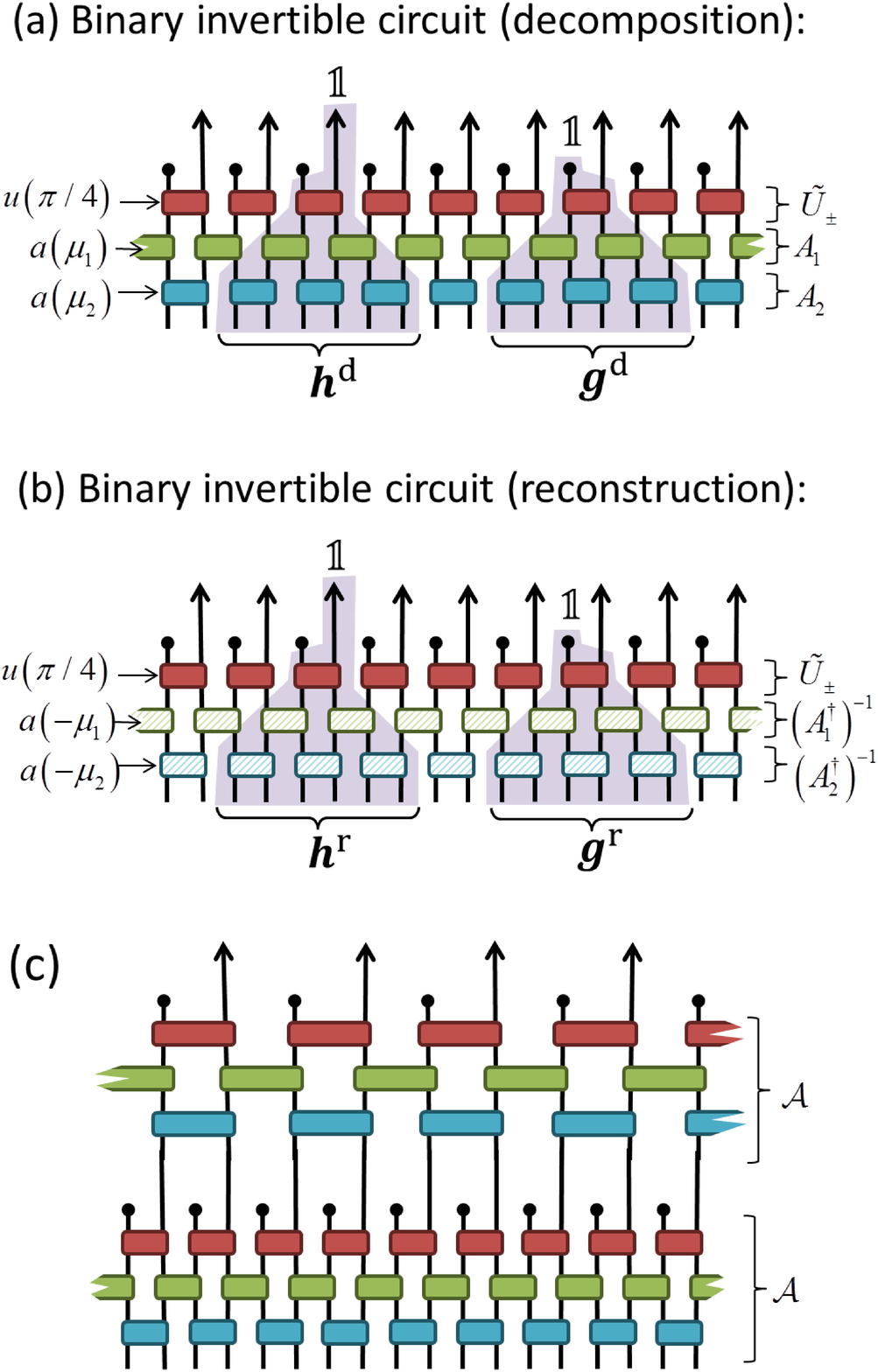}  \end{centering}
  \caption{(a) Decomposition binary invertible circuit $\mathcal A$ of depth $N=2$, see Eq. \ref{eq:s9e2}. Layers $A_k$ are composed of a direct sum of $2\times 2$ invertible symmetric matrices $a$ of Eq. \ref{eq:s9e1}, while layers $\tilde U_\pm$ are composed of two site unitary gates $u$. Transforming a unit vector $\mathbbm 1$ gives either the symmetric scaling sequence $\bm{h}^\textrm{d}$ or antisymmetric wavelet sequence $\bm{g}^\textrm{d}$. (b) Reconstruction binary invertible circuit $\left( {\mathcal A}^\dag \right)^{-1}$ for the circuit in (a). Reconstruction scaling $\bm{h}^\textrm{r}$ and wavelet $\bm{g}^\textrm{r}$ sequences are shown. (c) Multi-scale circuit formed from composition of the binary invertible circuit $\mathcal A$ of (a).} \label{fig:BiorthogCircuit}
\end{figure}

We consider a circuit with the same structure as the binary unitary circuits considered in Sect. \ref{sect:circuit}, but where the unitary constraint on the $2 \times 2$ matrices has been removed, and instead the matrices are constrained to be reflection symmetric. Let $a(\mu)$ be a $2\times 2$ reflection symmetric matrix parameterized by real parameter $\mu$,
\begin{equation}
a(\mu ) = \left[ {\begin{array}{*{20}{c}}
  1&\mu  \\ 
  \mu &1 
\end{array}} \right], \phantom{xxx} |\mu|\ne 1, \label{eq:s9e1}
\end{equation}
and let $\mathcal{A}$ be an invertible matrix of dimension $2M\times 2M$ for integer $M$. We say that $\mathcal{A}$ is a \emph{binary invertible circuit} of depth $N$ if it can be decomposed into sublayers,
\begin{equation}
\mathcal{A} = {A_N}{A_{N - 1}} \ldots {A_2}{A_1}{\tilde U_ \pm }, \label{eq:s9e2}
\end{equation}
see also Fig. \ref{fig:BiorthogCircuit}(a), where $\tilde U_ \pm$ is a direct sum of unitary gates $u(\pi/4)$ as defined in Eq. \ref{eq:s7e3}. Here $A_k$ is a $2M\times 2M$ matrix given as the direct sum of $M$ matrices $a(\mu)$, 
\begin{equation}
{A_k} =  \ldots \oplus {a\left( {{\mu _k}} \right)} \oplus {a\left( {{\mu _k}} \right)} \oplus {a\left( {{\mu _k}} \right)} \oplus \ldots, \label{eq:s9e3}
\end{equation}
and each sublayer $A_k$ is offset by one site from the preceding sublayer. Notice that the depth $N$ circuit $\mathcal{A}$ is parameterized by the $N$ real numbers $\{\mu_1, \mu_2, \ldots, \mu_N \}$ restricted such that $|\mu_k| \ne 1$.

We wish to interpret the binary invertible circuit $\mathcal{A}$ as the \emph{decomposition} part of a two-channel filter bank. Let $\bm{q}$ be a vector of length $2M$, which we transform to new vector $\bm{p}$ using the binary invertible circuit $\mathcal A$, 
\begin{equation}
\bm{q}^\dag \mathcal{A} = {\bm{p}^\dag }. \label{eq:s9e4}
\end{equation}
The original vector $\bm{q}$ can be reconstructed from $\bm{p}$ as,
\begin{equation}
{\left( {{\mathcal{A}^\dag }} \right)^{ - 1}}\bm{p} = \bm{q}, \label{eq:s9e5}
\end{equation}
where ${\left( {{\mathcal{A}^\dag }} \right)^{ - 1}}$, which we refer to as the dual of $\mathcal{A}$, is the \emph{corresponding} reconstruction part of a two-channel filter bank. 

Let us now examine the form of the dual circuit. The inverse-transpose can be taken of each sublayer individually,
\begin{equation}
{\left( {{\mathcal{A}^\dag }} \right)^{ - 1}} = \left( {A_N^{\dag}}\right)^{-1}  \ldots  \left( {A_2^{\dag}}\right)^{-1} { \left( {A_1^{\dag}}\right)^{-1}}{\tilde U_ \pm }, \label{eq:s9e6}
\end{equation}
where again each sublayer $\left(A_k^\dag\right)^{-1}$ is offset by one site from the preceding sublayer, see also Fig .\ref{fig:BiorthogCircuit}(b). Notice that the transpose of $\tilde U_ \pm$ it is own inverse. Similarly, since it can be shown that (for $|\mu|\ne 1$),
\begin{align}
  a{(\mu )^{ - 1}} & \propto \left[ {\begin{array}{*{20}{c}}
  1&{ - \mu } \\ 
  { - \mu }&1 
\end{array}} \right] \hfill \\
   & = a\left( { - \mu } \right), \label{eq:s9e7}
\end{align} 
it follows that $\left(A_k^\dag\right)^{-1}$ is a given as the direct sum of $M$ matrices $a(-\mu)$,
\begin{equation}
\left(A_k^\dag\right)^{-1} \propto  {a\left( {{-\mu _k}} \right)} \oplus {a\left( {{-\mu _k}} \right)} \oplus {a\left( {{-\mu _k}} \right)} \oplus \ldots, \label{eq:s9e8}
\end{equation} 

We label the (decomposition) scaling and wavelet sequences as  $\bm{h}^\textrm{d}$ and $\bm{g}^\textrm{d}$ respectively, which are given from $\left\langle {\mathcal{A} \times {\mathbbm{1}_r}} \right\rangle$, and the (reconstruction) scaling and wavelet sequences as $\bm{h}^\textrm{r}$ and $\bm{g}^\textrm{r}$ respectively, which are given from $\left\langle {\left(\mathcal{A^\dag}\right)^{-1} \times {\mathbbm{1}_r}} \right\rangle$, see also Fig. \ref{fig:BiorthogCircuit}(a-b). These coefficient sequences have the following properties:
\begin{itemize}
\item $\bm{h}^ \textrm{d}$ and $\bm{h}^ \textrm{r}$ are a edge-centered symmetric sequences of (even) length $2N+2$ elements.
\item $\bm{g}^ \textrm{d}$ and $\bm{g}^ \textrm{r}$ are a edge-centered antisymmetric sequences of (even) length $2N+2$ elements.
\end{itemize}
Once again the binary invertible circuit can be composed to form a multi-scale circuit,
\begin{equation} 
\mathcal{A} \circ \mathcal{A} \circ \mathcal{A} \circ \mathcal{A} \circ \ldots, 
\end{equation} 
see also Fig .\ref{fig:BiorthogCircuit}(c), which encodes a family of antisymmetric biorthogonal wavelets of dilation factor $m=2$.

%%%%%%%%%%%%%%%%%%%%%%%%%%%%%%%%%%%%%%%%%%%%%%%%%%%%%%
%%%%%%%%%%%%%%%%%%%%%%%%%%%%%%%%%%%%%%%%%%%%%%%%%%%%%%

\subsubsection{Example 6: edge-centered symmetric biorthogonal wavelets}
\begin{figure} [!htb]
  \begin{centering}
\includegraphics[width=8.5cm]{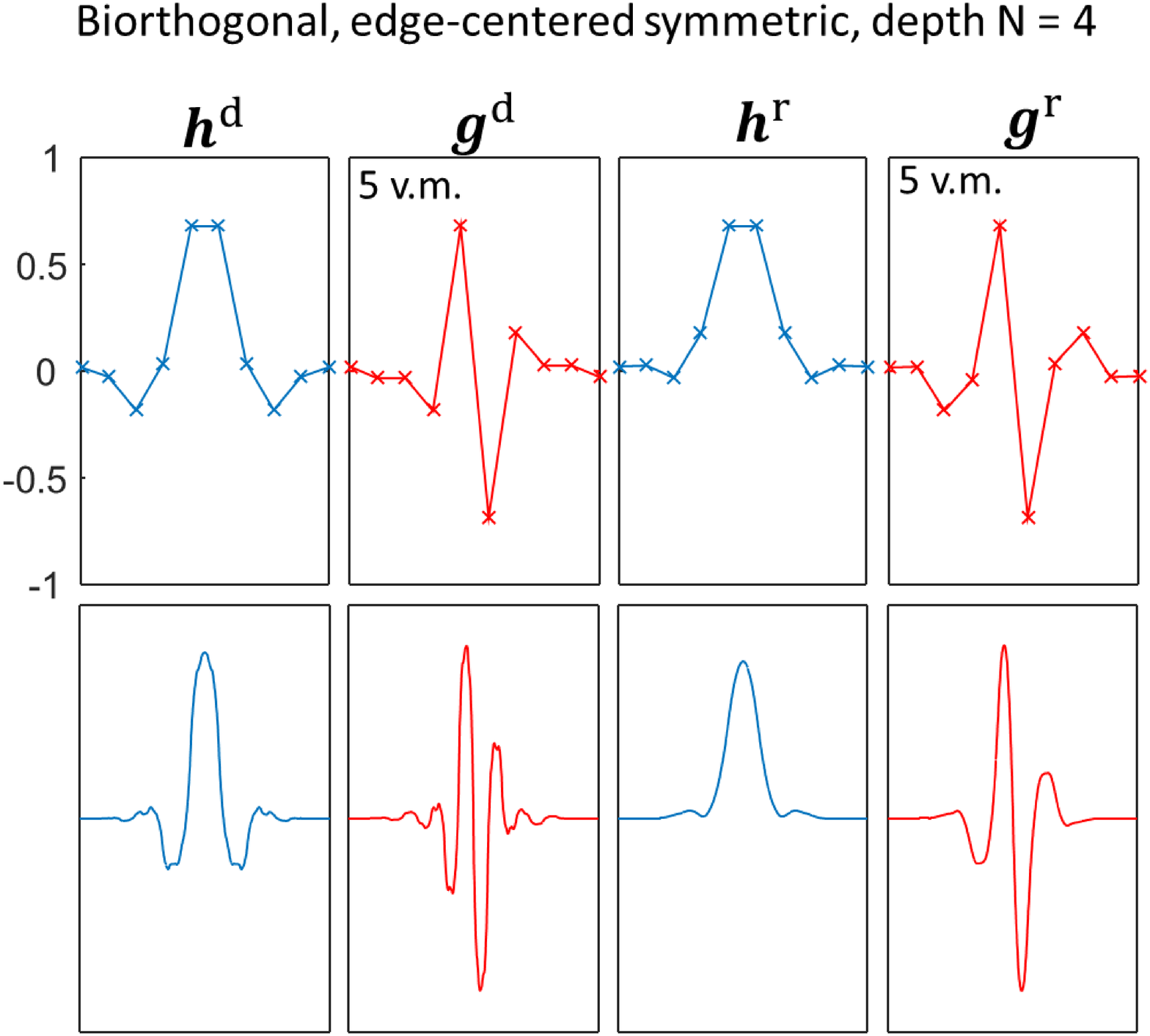}  \end{centering}
  \caption{(Top) Plots of the decomposition scaling $\bm{h}^\textrm{d}$ and wavelet $\bm{g}^\textrm{d}$ sequences, and reconstruction scaling $\bm{h}^\textrm{r}$ and wavelet $\bm{g}^\textrm{r}$ sequences, given from a depth $N=4$ binary invertible circuit with parameters $\mu_k$ as specified in Tab. \ref{table:biorthog}. Both wavelet sequences possess 5 vanishing moments (v.m.). (Bottom) The corresponding scaling functions and wavelets in the continuum limit. } \label{fig:BiorthogWave}
\end{figure}

Here we provide an example of biorthogonal wavelets constructed using a binary invertible circuit. For a depth $N=4$ binary invertible circuit we use algebraic methods to fix the parameters $\{\mu_1, \mu_2, \mu_3, \mu_4 \}$, see Tab. \ref{table:biorthog}, such that both the decomposition $\bm{g}^ \textrm{d}$ and reconstruction $\bm{g}^ \textrm{r}$ wavelets have vanishing moments for $\alpha = [0,1,2,3,4]$. The resulting antisymmetric wavelets are depicted in Fig. \ref{fig:BiorthogWave}.

%%%%%%%%%%%%%%%%%%%%%%%%%%%%%%%%%%%%%%%%%%%%%%%%%%%%%%
\begin{table}[htb]
\centering
\begin{tabular}{|c||c|c|}
\hline
       & $N = 4$              \\ \hline
$\mu_1$       & \phantom{x} -0.25684952118 \phantom{x} \\ 
$\mu_2$       &  \phantom{-}0.85058116979   \\ 
$\mu_3$       &  \phantom{-}0.05040158211  \\ 
$\mu_4$       & -0.83314630482  \\\hline
\end{tabular}
\caption{Parameters defining the depth $N=4$ binary invertible circuit, as depicted in Fig. \ref{fig:BiorthogCircuit}, which produce the wavelets depicted in Fig. \ref{fig:BiorthogWave}.} \label{table:biorthog}
\end{table}
%%%%%%%%%%%%%%%%%%%%%%%%%%%%%%%%%%%%%%%%%%%%%%%%%%%%%%

%%%%%%%%%%%%%%%%%%%%%%%%%%%%%%%%%%%%%%%%%%%%%%%%%%%%%%
%%%%%%%%%%%%%%%%%%%%%%%%%%%%%%%%%%%%%%%%%%%%%%%%%%%%%%

\section{Conclusions} \label{sect:conclusion}
The main objectives of this manuscript were to recast wavelets in the circuit formalism familiar to multi-scale methods used for simulating quantum many-body systems, thereby further elaborating on the connections between these fields, and also to explore the use of the circuit formalism in the design of novel wavelet transforms. 

%%%%%% circuit representation of Daubechies is useful
The representation of common orthogonal wavelet families, such as Daubechies wavelets, symlets and coiflets, as unitary circuits composed of $2\times 2$ rotation matrices, which are parameterized by rotation angles $\theta_k$, appears to be a very natural one. The circuit representation makes explicit key properties of the wavelets (such as orthogonality), facilitates in the understanding of aspects of their construction (such as the cascade algorithm), and is amenable to their efficient numeric implementation. Furthermore, as evidenced in Tabs. \ref{Tab:Daubechies}, \ref{Tab:Symlets} and \ref{Tab:Coiflets} and Fig. \ref{fig:Angles}, parameterization in terms of rotation angles affirms a clear-cut relation between different orders of wavelets within the same family.

%%%%%% novel wavelets
The families of wavelets given in Sect. \ref{sect:novel} were constructed via a two-step process that involved (i) the design of the circuit that builds desired properties into the parameterization (e.g. orthogonality, symmetry, dilation factor) and (ii) numerical optimization of the parameters according to some specified criteria (e.g. to achieve the maximal number of vanishing moments). Interestingly, these are the same two steps involved in the implementation of MERA for the study of quantum many-body systems, where the criteria for the numerical optimization in (ii) then involves minimizing the energy of the MERA with respect to a system Hamiltonian. In most of the examples presented in this manuscript the Nelder-Mead algorithm, a standard numeric algorithm for minimizing a multi-dimensional objective function, was used in the optimization of the circuit parameters, although more sophisticated numerical or analytic methods could, in general, also be employed. 

%%%%%%
The examples provided were intended to demonstrate the utility of the circuit formalism in wavelet design, but certainly do not constitute an exhaustive list of the wavelets that can be constructed using this approach. For each of the circuit examples provided, one could straight-forwardly consider higher order wavelets based on larger depth circuits, as well as alternative optimization criteria for the free parameters $\theta_k$ of the circuits. Many other circuit designs are also possible, including circuits in higher dimensions, analogous to higher dimensional implementations of MERA\cite{Alg2,MERAapp4}, which could represent novel families of non-separable orthogonal $2D$ (or $3D$) wavelets. Despite only intending to serve as examples, many of the wavelets given appear to perform well in practical applications. Though not presented in this manuscript, preliminary investigations suggest that the wavelets described in examples 2, 3 and 6 perform similarly to, or in some cases exceed the performance of, the CDF-9/7 wavelets in application to image compression. The use of circuits in optimal wavelet design for specific applications remains an interesting direction for future research.  

The authors acknowledge support by the Simons Foundation (Many Electron Collaboration). SRW acknowledges funding from the NSF under grant DMR-1505406.

\end{document}